# Cooling performance of a wick consisting of closely packed rods at moderately high heat loads

Navneet Kumar[1#], V S Jasvanth[2], Amrit Ambirajan[2], and Jaywant H Arakeri[1]

[1]*Department of Mechanical Engineering, Indian Institute of Science, Bangalore – 560012, India*
[2]*Experimental Heat Transfer Section, Indian Space Research Organisation, Bangalore – 560012, India*

*Abstract* − We propose a new class of wicks, consisting of closely packed circular rods, whose evaporative capacities have been measured at different heat loads ranging between 0.05W/cm$^2$ to 8W/cm$^2$. The experiments were performed for two liquids, water and highly volatile pentane, in a specially designed setup to understand transient and steady state cooling characteristics of the proposed wicks. Heat interception and vapour release occur on the same side in these experiments. These wicks released ~50% of the supplied heat load as the latent heat; this value remained nearly constant between heat loads of 0.5 and 8W/cm$^2$. These wicks have the unique characteristic of potentially very high and rapid capillary rise induced by near-zero radii (NZR) of contacts formed between the rods in contact; liquid region reaching the end in NZR has been called 'corner' meniscus. While the bulk liquid (present between three rods) may recede, depending on the heat load, the corner meniscus remains pinned; this unique feature thus leads to sustained high evaporation rate requirements. This remarkable characteristic seems advantageous compared to a regular wick, whose cooling performance depends on the heat loads. We also performed 3-D unsteady state numerical simulations to understand the effect of rod diameter and materials' thermal conductivity on the overall wicks' performance. Steady state temperature value was in fairly good agreement with the ones measured in experiments. Based on the geometry of the corner film, fluid mechanics of liquid transport, and the heat transfer aspects, we present a design of suitable wicks as per the requirement. These new configurations can represent a separate class of wicks and may replace the regular wicks in current and futuristic cooling devices.

## 1. Introduction

Heat pipes (HP) [1,2] are passive devices, which efficiently transfer heat from a hot spot to another location. This ability of HP has led to its use in many electronic devices; one such use is in spacecraft during its re-entry into the atmosphere. Phase change is the main mechanism by which high cooling rates are obtained. In the majority of these systems, heat is conducted, through the device casing or wall, from the hot spot to the evaporator section (Figure 1a), consisting of a wick (a porous medium) holding the evaporating liquid of an HP. Wicks in a regular HP is made up of spheres, sometimes sintered [3]. The liquid evaporates and the vapours thus generated are driven, by the pressure gradient, to a separate section, a condenser, where they condense back into the liquid (Figure 1a) thereby releasing the latent heat. The condensed liquid comes back to the evaporator section via the capillary action through the wick. In loop heat pipes (LHP) [4,5] and capillary pumped loops (CPL), the wick is limited to the evaporator section, unlike HP, where the wick runs throughout the length of HP. Since the entire process is pressure-driven, it is important for the interfacial force acting on the liquid-vapour (LV) menisci to be higher than the total pressure loss of the system [3,6]; this eventually decides the pore size required for the (conventional) wick. Before discussing the salient features of the proposed wick, we must, first, discuss the conventional wicks, their selection, working, and limitations.

*Wick selection and types*

In general, for any porous medium consisting of a packed bed of solids (such as the one in Figure 1b), we may write

$$h \propto \left(\frac{1}{r}\right) \qquad (1)$$

*Corresponding author at: Fluid Mechanics Laboratory, Mechanical Engineering, IISc Bangalore, 560012, India.
E-mail address: *navneet01011987@gmail.com*, *navneetkumar@iisc.ac.in*

$$\kappa \propto \frac{\phi^3}{(1-\phi)^2} r^2 \qquad (2)$$

where $h[m]$ is the height-rise of the liquid in the wick, $r[m]$ is the mean pore size of the wick, $\kappa[m^2]$ is the absolute permeability, and $\phi$ is the porosity of the medium. Eq. (1) seems to hold as long as the packed bed is made up of (regular or irregular) particles.

It was shown [7] that, during evaporation from a conventional porous medium, three length scales are important, each corresponding to competition between two of the three forces (surface tension, gravitational, and viscous). It was derived that, at low (0-10 mm/day) and moderate (10-100 mm/day) rates of evaporation, the viscous forces can be neglected and the only relevant length scale is gravity-based, defined as

$$L_g = \frac{2\sigma}{g(\rho_l - \rho_g)}\left(\frac{1}{r_1} - \frac{1}{r_2}\right) \qquad (3)$$

Where, $\sigma[N/m]$ is the interfacial surface tension at the LV meniscus, $\rho_l$ and $\rho_g[kg/m^3]$ are the evaporating liquid and invading gaseous phase densities, respectively, $g[m/s^2]$ is the gravitational acceleration, $r_1$ and $r_2[m]$ are the radii of the smallest and the largest pores in the system. It was advised that $r_1$ and $r_2$ be determined using water retention curves and not directly from the particle sizes.

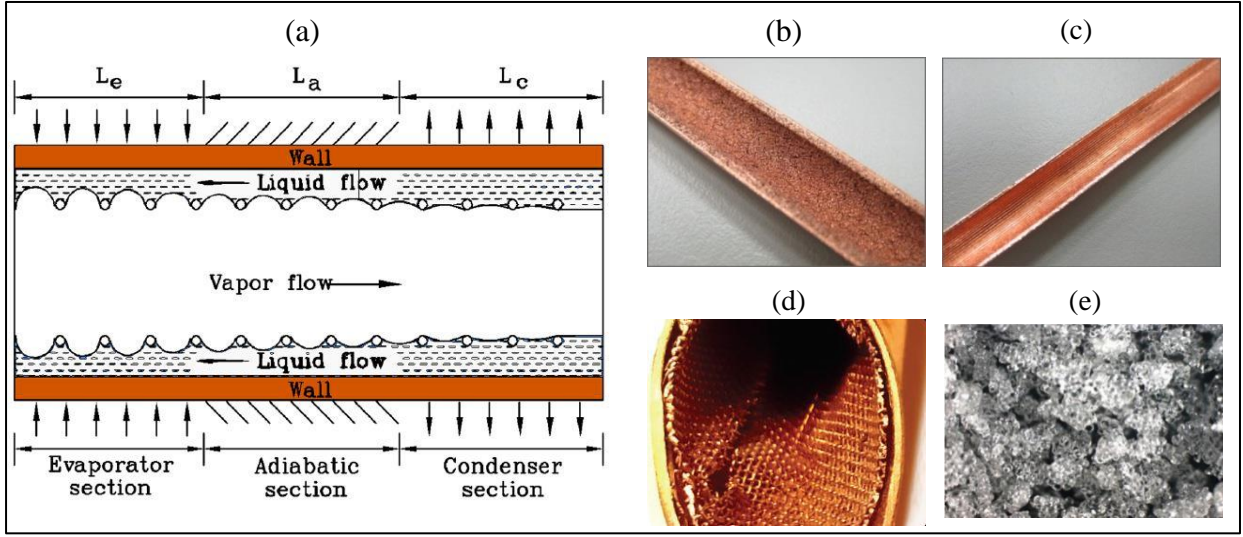

Figure 1 (a) Axial cross-section of a heat pipe (taken from [8]) and (b-e) different types of wicks (Photo courtesy: Google Images) being used.

In selecting a wick, as per our knowledge, four parameters are important: (a) wicking height, (b) permeability, (c) particle (or pore) size distribution, and (d) evaporation rate (or heating load) requirements. The other key parameter, which has been greatly underestimated, is the rate at which liquid rises in a wick, i.e. $dh/dt$. We will discuss this transient property of the wick soon. We see that a lower pore size offers higher capillarity (Eq. 1) but diminishes the liquid flow velocity in the wick since the permeability is greatly reduced (Eq. 2). LHP with regular wicks, thus, must maintain a balance between the desired high interfacial forces and high permeability. A cooling device, with a fixed evaporating liquid, has a limited working range of heat loads, which is eventually governed by the mean pore size of the wick. Note that the particle size distribution can be avoided using (nearly) mono-disperse particles [9]. Properties of porous systems with textural contrasts have also been previously explored.

Large interfacial force and higher permeability can be achieved by the use of a bi-porous wick [10], consisting of two different, yet intermixed, sizes of particles (Figure 1e). Another configuration could be to use two different particle sizes, without mixing, either in vertical [11,12] or horizontal [12-14] layers or in annular configuration [15]. Higher permeability is achieved via the larger particles, while high capillarity is supported by the smaller ones. Depending on the combinations of such sizes a whole class of wicks could be configured. However, higher pressure and viscous losses are of major concern here. In groove-based wicks (Figure 1c), wicking occurs via machined corners such as in a square or rectangular capillary. The corners in these groove-based wicks, however, have a finite degree of roundedness [16], which limits its capillarity; this is, therefore, a fabrication and machinability issue. Performance of these wicks at high heat loads remains unclear compared to the regular wicks (Figure 1b,d).

*Limitations of regular wicks*

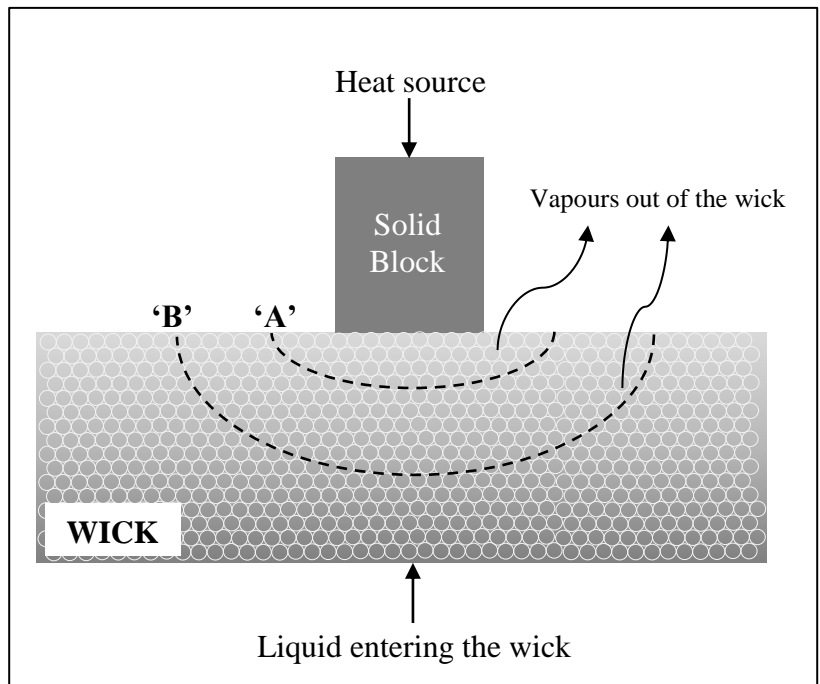

Figure 2 Heat and mass flow in an evaporator in LHP, inspired by [6]. 'A' and 'B' show (hypothetically) the steady state positions of the liquid-vapour meniscus (dashed line), in the wick, at 'low' and 'high' heat loads, respectively.

Of particular interest are LHPs and CPLs; unlike HPs, where liquid transport in the wick and from the condenser to the evaporator zone is driven by the surface tension, here the liquid movement from the condenser region to the wick is driven by the developed pressure gradient. According to the schematic seen in Figure 2, a solid block (of high thermal conductivity) lies between the wick of an LHP/CPL and the heating source. For easier phase change, the LV menisci are desired to be positioned as near as possible to the heat source. Such a meniscus is marked 'A' in Figure 2. Note that the wick above meniscus 'A' is fully dry and is fully wet below it. A portion of the conducted heat is used in the phase change of the liquid and consequently, the evaporated vapours move out of the wick as seen in Figure 2. As some of the liquid is consumed, during evaporation, the wick tries to suck in the required liquid amount, to keep up with the evaporation rate. Note that in LHPs and CPLS, the (primary) wick is also hydraulically connected to either a secondary wick or some other source. This balance between the lost (evaporated) liquid and its immediate recovery in the wick is needed for sustained evaporation. At a higher heat load, the system achieves a separate steady state where the LV meniscus is situated away from the heat source ('B' in Figure 2). This new position comes at a great cost: increased operating temperature and reduced cooling capacity. The *additional* resistance, offered by the drier wick portion, increases its temperature, which is undesirable. Additionally, the evaporated vapours now have to travel more through the tortuous paths (voids between the spheres) within the dry part of the wick. With increasing heat loads, the LV menisci recede even deeper into the wick [17] and eventually '*dry-out*'; LHPs/CPLs fail at this point.

The objective of the present work is to test the cooling performance of a new type of wick, consisting of closely packed circular rods, at various heat loads. Some properties of evaporation from a stack of rods have been previously reported [18,19]. The most important feature of this wick is the presence of the '*near-zero radii* (NZR)' contacts throughout the wicks' length. These NZR of contacts maintain the LV interface near the evaporating (heating region) end of the wick thus sustaining high cooling rates. The other interesting feature of this wick is the non-tortuous pathways for the vapours to escape, unlike highly tortuous ones in case of the regular wicks. In short, the proposed unconventional wick offers three advantages,

1. Much higher liquid rise due to NZR of contacts formed between touching rods, and
2. Non-tortuous straight pathways for the vapours to escape, and
3. A high rate of liquid rise.

The paper is arranged as follows. In Section 2, we discuss the geometrical features of the rods in contact, especially of the corner film and the liquid velocity along it. The materials selected as the proposed wick and the method adopted to characterize its cooling performance, through a unique experimental strategy, has been detailed in Section 3. In section 4, we show results of the present experimental study along with a thorough heat budget and results from the simulations. We conclude this study in Section 5. In the Appendices (A-D), we have shown the effect of varying rods' diameter, rods' length, separated (not in contact) rods; these additional details would be useful in wick selection depending on the need.

## 2. Theoretical considerations

In previous studies [18,19], the primary objective was to understand the evaporation characteristics from a stack of packed rods at low heat loads. The number of rods in these studies ranged from ~300 to ~5000. It was also shown that a large number of rods, stacked in a container, lead to anisotropy (unwanted and unavoidable gaps at few places), which may reduce the capillary rise [20,21]. Here, we first discuss the geometrical aspects of the corner film formed in the NZR of contacts between two rods.

*Corner film parameters*

Figure 3 shows the top and front views of the corner film between two rods in contact. This configuration can be achieved in two different ways (a) when the rods are dipped in a pool of liquid (Figure 3c) and (b) during evaporation experiments (Figure 3d). The dotted line in Figure 3b represents the free liquid level or the position of the bulk meniscus during evaporation. Note that ideally, the meniscus in the corner would reach the top end of the rods [20]. The input parameters are: rod diameter ($R_{rod}$) and separation between the rods ($d$), which can be deliberately put or is governed by the flatness and roughness of the rods.

For simplicity, we assumed the contact angle, between the LV meniscus and the rod, to be zero. A non-zero contact angle can be implemented easily. At any vertical position, as per Figure 3, the corner film can be characterized by the following geometrical parameters,

Radius of curvature ($r_Z$) : Balancing the hydrostatic pressure and the surface tension yields,

$$\frac{\sigma}{r_Z} = \rho_l g Z \Rightarrow r_Z = \frac{\sigma}{\rho_l g Z} \tag{4}$$

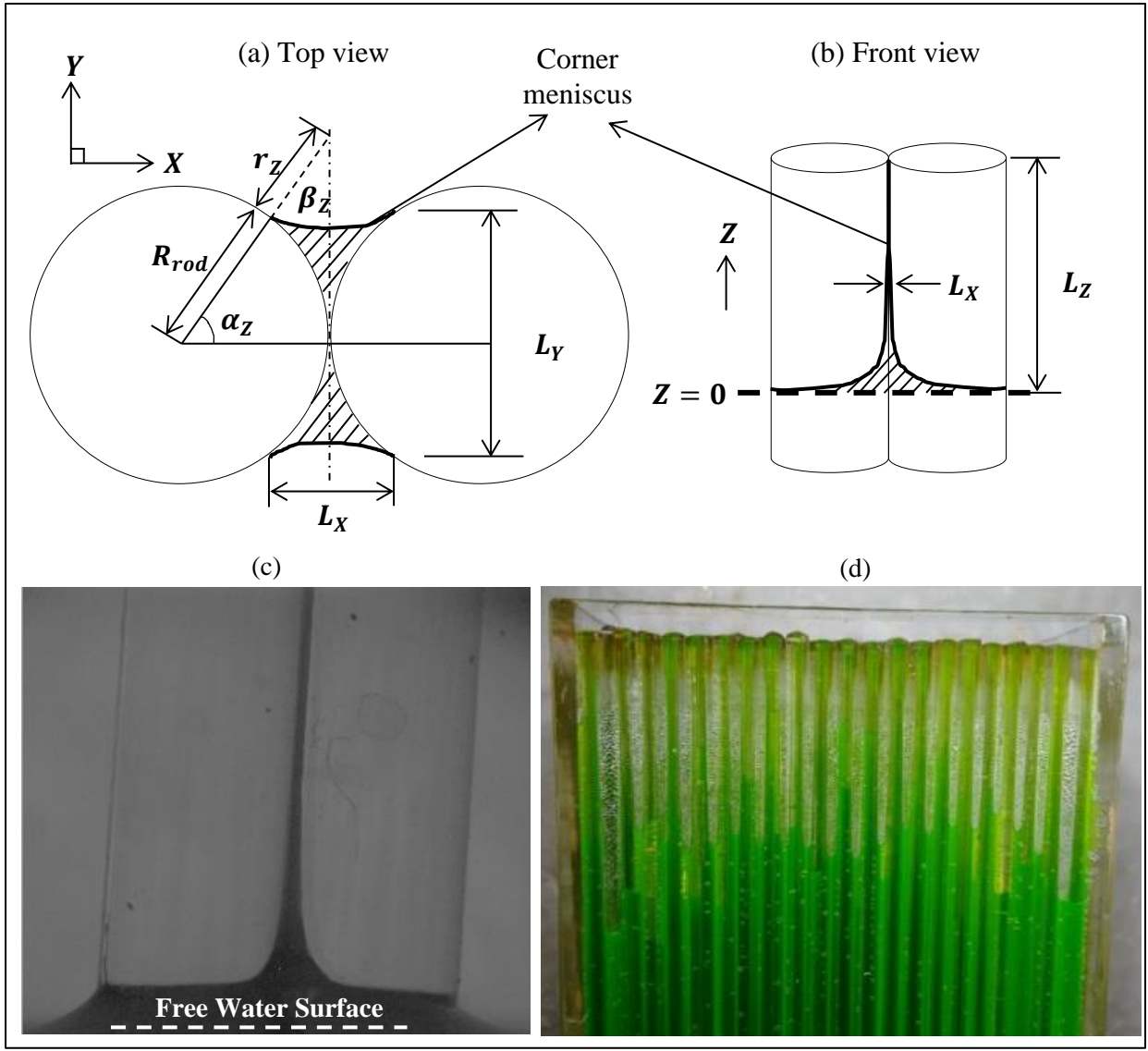

Figure 3 Geometry of a corner film formed between two vertical rods in contact. (a) & (b) show the top and the front views of the film, respectively at any Z location. Z is positive in the vertically upward direction and gravity is acting vertically downwards. A corner meniscus formed in the NZR of contact between two 3 mm diameter glass rods [18] when they were dipped in water. Green pillars in (d) show the corner films, in the evaporation experiment, between rods and the wall that reach the top of the rods. The green colour is due to the fluorescein dye mixed in water. Drops (seen after magnifying the image) are the condensed water vapour formed in the space between two rods and the wall.

Angle $(\beta_Z)$ : $\beta_Z = \sin^{-1}\left(\frac{R_{rod}}{R_{rod}+r_Z}\right) = \sin^{-1}\left(\frac{R_{rod}}{R_{rod}+\left(\frac{\sigma}{\rho_l g Z}\right)}\right)$ (5)

Angle $(\alpha_Z)$ : $\alpha_Z = \frac{\pi}{2} - \beta_Z = \cos^{-1}\left(\frac{R_{rod}}{R_{rod}+\left(\frac{\sigma}{\rho_l g Z}\right)}\right)$ (6)

Film width $(L_X)$ : Maximum width of the film at any Z-plane (see Figure 3a).

$$L_X = 2R_{rod}(1 - \sin\beta_Z) = \frac{2R_{rod}\, r_Z}{R_{rod} + r_Z} = 2r_z\left(1 + \frac{r_z}{R_{rod}}\right)^{-1} \quad (7)$$

Transverse film width $(L_Y)$ : Maximum width of the film (along Y direction).

$$L_Y = 2R_{rod}\cos\beta_Z = \frac{2R_{rod}\sqrt{r_Z^2 + 2R_{rod}\, r_Z}}{(R_{rod} + r_Z)} \quad (8)$$

In the limit $\frac{r_Z}{R_{rod}} \ll 1$ or $\frac{Z}{R_{rod}} \gg 1$ we get,

$$L_X \propto Z^{-1} \text{ and } L_Y \propto Z^{-0.5} \quad (9)$$

Film length $(L_Z)$ : Distance between the liquid free surface and the top end of the corner film.

Note that the radius of curvature in the other plane (perpendicular to $r_Z$) has been neglected. We now briefly discuss the variations of the corner film properties. We fix the rod length to 75 mm (the ones used in the experiments) and study two values of rod diameters (1 and 5 mm). Note that, in LHPs and CPLs, the conventional sintered spheres-based wicks are moulded in the form of a cylinder; these are ~50 mm long and 10-20 mm in diameter. Finally, we consider two evaporating liquids, water and n-pentane.

Figure 4 shows the variation of the four geometrical parameters of the corner film versus the vertical extent $(Z)$; $Z = 0$ is considered as the position of the free liquid surface. The axes in sub-figures (a, c, and d) are logarithmic, while they are linear in sub-figure (b). Note that all the length scales are mentioned in 'mm'. The radius of curvature of the corner meniscus $(r_Z)$ depends on the liquid density and the interfacial surface tension but not on the rod diameter; this is evident from Eq. (4). On log-log scales $r_Z - Z$ curves are linear, as seen in Figure 4a, with slope equal to $\sigma/\rho_l g$. The value of $r_Z$ reduces from ~1000 mm at $Z \to 0$ to $< 1$ mm at $Z = 10$ mm. The film widths $L_X$ and $L_Y$ both decreases non-linearly and drastically when $Z > 1$ mm. At higher $Z$ values, $L_X$ follows Eq. (8) as seen in Figure 4c. Note that the formulae used for defining the corner film geometry cannot be used at low values of Z or when $Z \to 0$. In the case of three rods in contact, the corner film geometry does not change but an extra parameter, bulk meniscus, should be taken into account.

Apart from the length scales, a parameter of more significance is the half angle $\alpha_Z$. In cases of rods with larger diameters, $\alpha_Z$ reduces quickly (Figure 4b) with $Z$. Consider a unit cell (regular hexagon) as seen in Figure 5. The central rod 'R1' forms six corner films (CF1-CF6) and other rods form six (CF7-CF12) more. Note that six central films form when $\alpha_Z < 30°$ but when $\alpha_Z \geq 30°$, the films would start interacting with one another. Consider an example. For a system consisting of 5 mm diameter rods and pentane as the evaporative liquid, $\alpha_Z = 30°$ at $Z \sim 6.8$ mm (Figure 4b). Below this height, the corner films cannot be differentiated from the bulk meniscus. However, if the rod length is increased to 75 mm (ones used in this study), $\alpha_Z$ at the rods' top reduces to ~9.5⁰. Similarly, systems can be configured based on whether the corner film interaction is needed or not.

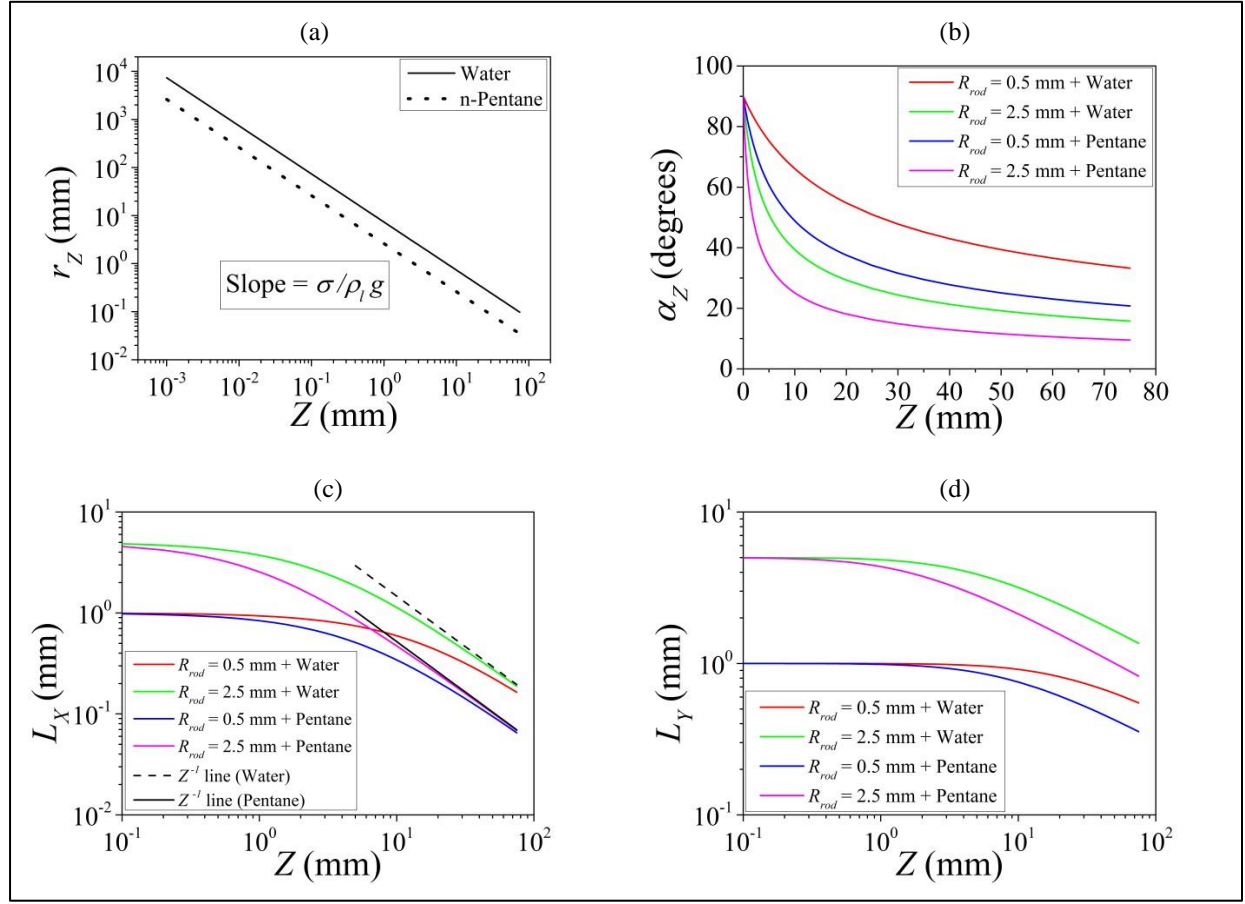

Figure 4 Variation with Z (vertical height) of various corner film geometrical parameters. The axes in (b) are normal, while in others they are logarithmic.

*Liquid velocity in the corner film*

Since it is the competition between the two rates, liquid removal rate near the heat source and the subsequent liquid recovery at the evaporating LV meniscus, which decides a wicks' performance, it is important to discuss the velocity scales during the transient rise of liquid in the NZR of contact. Three phases of liquid rise have been identified [22-26] during any capillary rise phenomenon, with or without the gravitational forces. These are (a) an initial phase, (b) a viscous phase, and (c) an inertial phase. In the initial phase, a meniscus forms where the contact angle needs to be changed from $90^0$ (flat free meniscus) to some finite value (for hydrophilic systems). Recently a universal scaling was proposed [26] for the transient capillary rise height, $h(t)$ in any continuous corner; these include between two rods and between two inclined planes. They validated their universal scale with a wide range of experiments. We directly use the result of [26] for $h(t)$, which is given as

$$h(t) \sim \left(\frac{\sigma^2 t}{\mu_l \rho_l g}\right)^{1/3} \qquad (10)$$

where $t$[s] is the elapsed time and $\mu_l$[kg/m − s] is the dynamic viscosity of the liquid. It was suggested [26,27] that Eq. (10) should only be used after an initial time ($t_i$), given by

$$t_i \sim \frac{10^3 \mu}{\sqrt{\sigma \rho_l g}} \tag{11}$$

Surprisingly, neither Eq. (10) nor Eq. (11) depends on the geometrical properties of the rod. At room temperature, we calculated $t_i$, for water and pentane, to be ~0.1 and 1 second, respectively. The corner meniscus is formed rapidly for commonly used evaporating liquids in LHPs and CPLs.

The other important velocity scale is the liquid flow velocity once the process (phase change at one end followed by simultaneous liquid recovery) becomes continuous. This velocity scale is determined by the steady state pressure gradient developed within the capillary film. Due to the liquid flow, we have an additional viscous pressure loss. Discussion on these parameters is detailed in Appendix D.

*Rate of the liquid rise in NZR of contact between two rods*

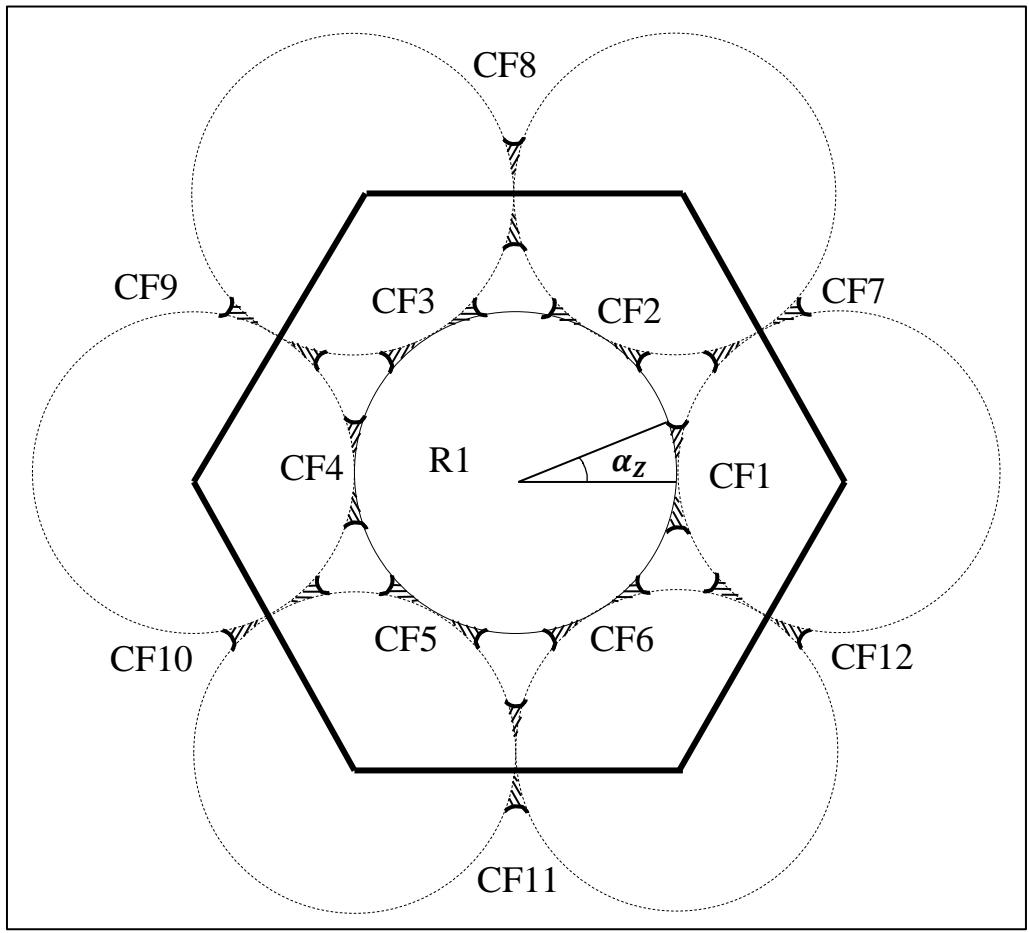

Figure 5 Schematic of a unit cell of a stack of rods. The central rod is in contact with six other rods with the same diameter. The hexagon (a unit cell for such systems) can be considered as the equivalent area for calculating the evaporation rate from such a system. The total number of films (shown as hatched) in this hexagon is nine (6 full and 6 half films). The notation 'R' and 'CF' denote the rod and corner film, respectively.

As discussed earlier, the liquid recovery rate or rate of liquid rise $\left(\frac{dh}{dt}\right)$ in a wick is more important than the total height rise ($h$). In case of the rods in contact ($d \to 0$), $h$ is limitless [20] anyways. The time derivative of $h$ can be considered as the liquid film velocity ($v_f$). We differentiate Eq. (10) to get

$$v_f = \frac{dh}{dt} \sim \frac{1}{3}\left(\frac{\sigma^2}{\mu_l \rho_l g}\right)^{1/3} t^{-2/3} \qquad (12)$$

Eq. (12) can also be written explicitly as a function of $Z$ and the fluid properties

$$v_f = \frac{\sigma^2}{3\mu_l \rho_l g}\left(\frac{1}{Z^2}\right) \qquad (13)$$

For pentane, $v_f$ is 57.9, 3.6, and 1.2 cm/s at $Z = 10$, 40, and 70 mm, respectively irrespective of the rods' diameter (see Table A1 in Appendix A). To sustain the evaporative demand, the following condition must be satisfied near the rods' top

$$E_l A_{ev} \leq v_f A_f \qquad (14)$$

Where, $E_l$[m/s] is the liquid removal rate over area $A_{ev}$[m$^2$] and $A_f$ is the total planar area occupied by the films. In case of a single corner film, its area $a_f$ can be simply approximated as

$$a_f = L_X L_Y \qquad (15)$$

We can further estimate $A_{ev}$ for a hexagonal unit cell (Figure 5) as

$$A_{ev} = 6\sqrt{3}\, R_{rod}^2 \qquad (16)$$

The total film area in the hexagon can be written as

$$A_f = (\text{number of films}) a_f = 9 a_f \qquad (17)$$

Note that the total number of single developed corner films in the hexagon is nine; six corresponding to the central rod and six half films (Figure 5). We may now rewrite Eq. (14) as

$$2 E_l R_{rod}^2 = \sqrt{3} v_f L_X L_Y \qquad (18)$$

In terms of rods' geometrical properties, we can write the maximum sustainable value of the rate of evaporation (for the system seen in Figure 5) as

$$E_l \sim \frac{2}{\sqrt{3}} \sin \alpha_Z (1 - \cos \alpha_Z)\left(\frac{\sigma^2}{\mu_l \rho_l g}\right)^{1/3} t^{-2/3} \qquad (19)$$

Note that Eq. (15) is used if $\alpha_Z$ is small, probably less than $10^0$. We can, therefore, use Eq. (19) to directly estimate the maximum permissible evaporation rate (or the maximum heat load indirectly) on the basis of the corner film parameters, evaporating liquid properties, and time. For pentane and 5 mm rod diameter, $E_l$ is ~8, 0.1, and 0.01 cm/s at $Z = 10$, 40, and 70 mm, respectively. For the same $Z$ values, $E_l$ increases to ~52, 0.7, and 0.1 cm/s when 1 mm diameter rods are considered (see Table A1 and A2 in Appendix A). These values suggest that at very high heat loads, we should either keep the bulk meniscus near the rods' top or use shorter rods. Figure 6 shows two different situations one each corresponding to just before and just after the maximum permissible evaporation rate. When the heat load is suddenly changed, the liquid velocity cannot match the evaporation demand and hence the interface leaves the rods' top (see Figure 6b); this situation is not desirable in cooling systems. Note that upon reducing the heat load, the interface would regain its position at the rods' top if the evaporation requirement is below the permissible limit. Further, Eq. (10) and Eq. (12) should be used only when the rods are in contact. For a finite gap (we can call this configuration as a separated corner) Eq. (10) is expected to change, the magnitude of which should be governed by $d$. For example: in the case of 5 mm diameter rods and pentane, the liquid rises to < 1mm when $d/R_{rod} = 1$ but if $d/R_{rod} = 0.02$ the liquid rises to ~8cm. More discussion on this parameter follows in Appendix B.

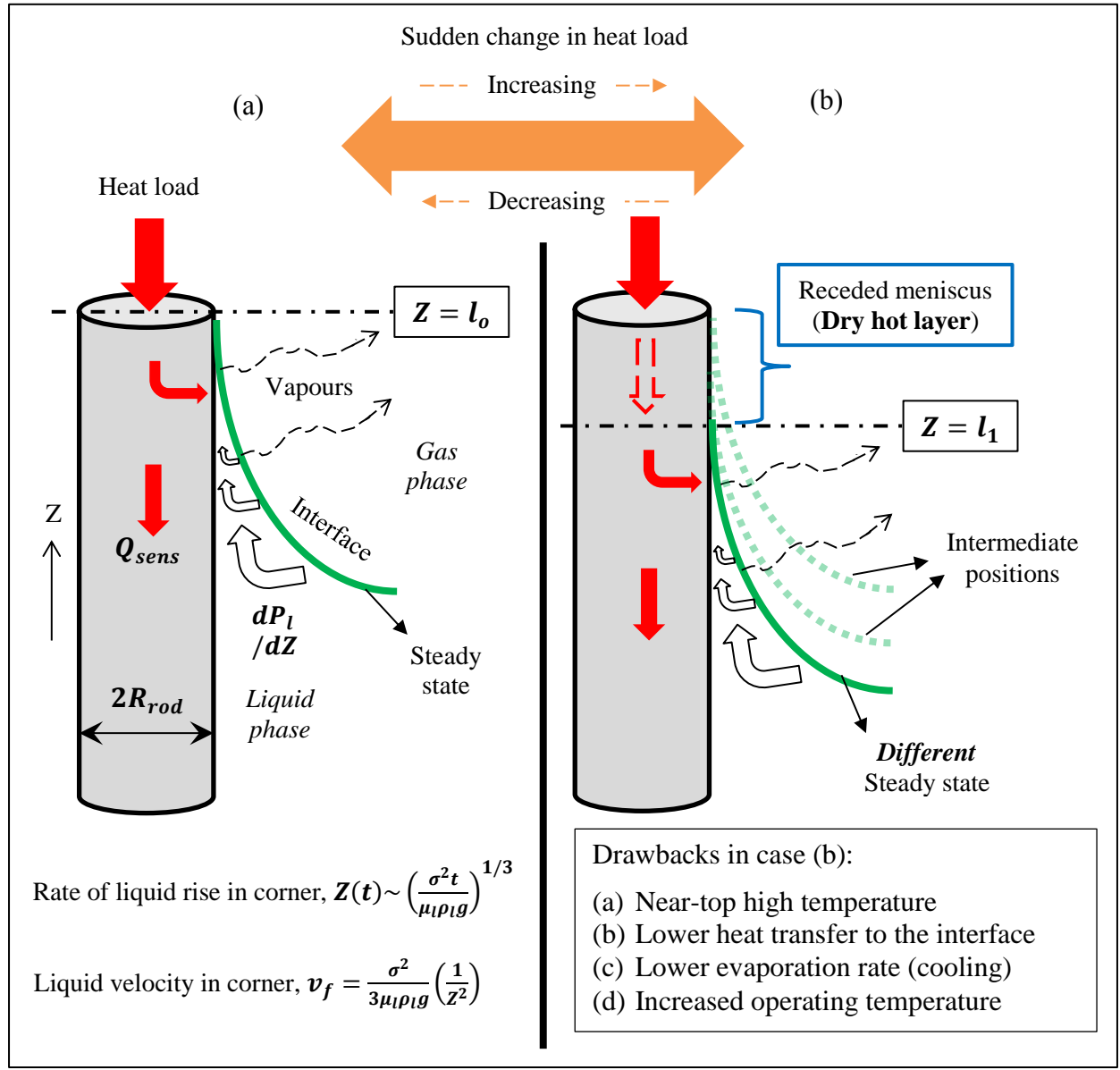

Figure 6 Schematics showing the conditions (a) just before and (b) just after the maximum permissible heat load limit. In case (a), the corner pulled liquid to the rods' top. Immediately after the permissible limit, the liquid velocity in the corner is unable to compete against the required evaporation rate (or the applied heat load); this forces the meniscus to recede away from the rods' top. A new steady state position is established in case (b) where the liquid velocity and the evaporation rate are again equal; however, this situation should be avoided.

## 3. Materials and methods

Experiments were designed to test the cooling capacity of the wicks consisting of the closely packed rods. Ideally, in a passive device, such as an LHP, the total mass loss in a cycle is zero; a weighing scale is hence useless. Instead, in the laboratory, we test the evaporative capacity of the wicks by measuring the amount of mass loss from them. For this purpose, two different methods of heating were adopted as mentioned below.

We briefly discuss the results and experimental method from a previous study [18]. Two important observations in these studies were (a) related to a characteristic evaporative length scale and (b) observation of long corner films. Based on these results, the experiments were designed in this investigation. In previous studies [9,12,18,19], external heat was provided, using an infrared heater, radiatively to obtain low heat loads (~0.1W/cm$^2$). The focus in these studies was on the transient behaviour of the wicks. The wicks were tested to study the effect of (a) rod diameter, (b) material, (c) heat loads, and (d) liquids (water and n-pentane). In similar conditions, wicks consisting of packed mono-disperse spheres were also tested previously [9]. The capillary film length, $L_{cap}$, (a measure of the effective depth up to which a wick can sustain high evaporation rates, viz., stage 1, in a porous medium) decreased with increasing spheres size [9,28]. It was further observed [29] that $L_{cap}$ decreased at higher heat loads when fixed spheres sizes were used. Another significant observation was a lower value of $L_{cap}$ when n-pentane, instead of water, was used [9]. Remarkably, in the limit of the experimental and geometrical parameters, $L_{cap}$ for wicks consisting of packed rods did not depend [18] either on the rods diameter (spanning diameter from 0.7 mm to 3 mm) or the liquid used or the heat load.

It was shown, theoretically, that the corner meniscus would reach infinity [20] if two rods are in contact. Figure 3c shows two 3 mm diameter glass rods in contact, a free water surface, and a capillary film (black vertical line) formed along the NZR of contact. The film width ($L_X$) reduces quickly (compared to its value at the free water level) and becomes nearly a constant as we go up; the change in $L_X$ is non-noticeable after some distance from the water level. This feature will be discussed in detail in the next section. In the case of three rods or two rods and a wall in mutual contact, along with the formation of corner menisci, a bulk meniscus (occupying the central space) also forms [21]. Similar water films were observed during the evaporation experiment as seen in Figure 3d; water was coloured using fluorescein dye. Green pillars represent water films formed between the glass rod and the acrylic wall [18,21]; the experiment was initiated when the rods were fully wet till the top. These green pillars exist nearly till the end of the experiment and are expected to exist longer for rods with larger lengths.

A separate experiment was designed to test the wick at moderately high heat loads as seen in Figure 7; different components of the experimental setup are numbered in Figure 7a. A strip heater ('1'), 25 x 25 mm in size (and 50W/220V rating), supplies the heat through conduction to the top part of the rods (5 mm diameter stainless steel (SS), 75 mm long). A silicone-based thermal interface material (TIM) was placed between the heater and the rods to maintain good contact. The rods are inserted in a brass tank (74 mm internal diameter, 18 mm depth, and 4 mm thick) that contains the liquid whose level was maintained via a siphon tube ('4') from a source tank ('2'), 100 x 100 x 50 mm in size. A switch ('3') was used to control the flow rate from the source to the brass tank. Mass loss was measured using a weighing scale ('5') with a precision of 0.01g. A schematic (Figure 7b) of the experimental setup shows these components. The highest heat load achieved was 8W/cm$^2$ and the lower values were obtained using a voltage regulator.

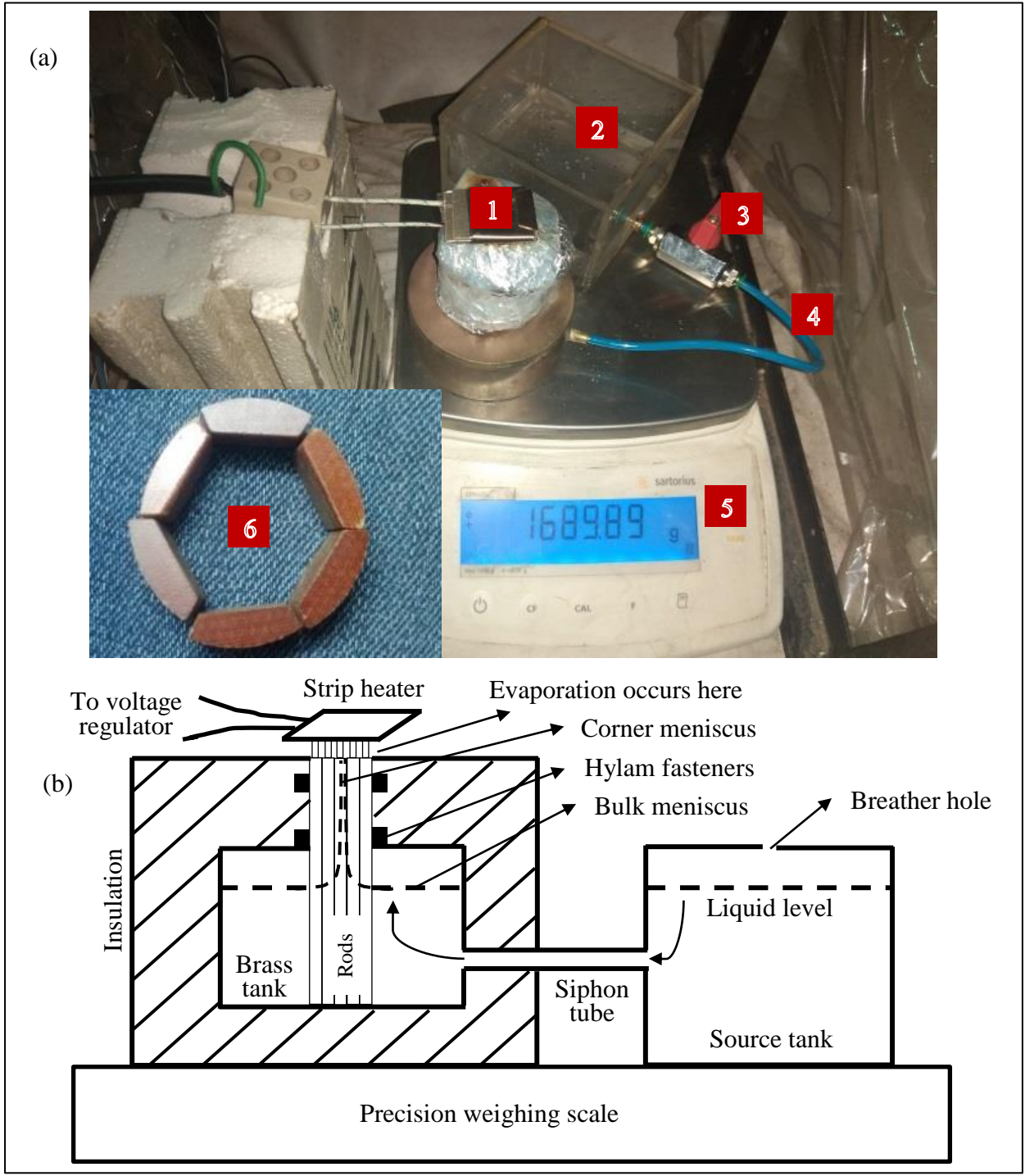

Figure 7 Experimental setup (a) for testing the wick characteristics at moderately high heat loads. Different components are numbered. The schematic of the experiment is shown in (b). The source tank has a tiny breather hole, which ensures atmospheric pressure at the liquid surface.

In order to avoid defects (anisotropy) when large numbers of rods are used, we used a total of 19 rods: 1 at the centre, 6 around it, and 12 in the last layer (see Figure 8a). Note that if we connect the centres of the outermost rods we get a regular hexagon. Accordingly, a hexagonal opening was etched out of the top face of the brass tank such that the rods could fit accurately.

The rods were held tightly (assuring the near-zero radii of contacts) using *Hylam fasteners* (numbered '6' in Figure 7) at two locations; one at the base of the brass tank and the other 35 mm above it. The Hylam (ability to withstand high temperatures) fasteners have a flat inner surface, which is tangent to the outermost rods while the outer surface is an arc of 16 mm radius. A circular clamp of the same (16 mm radius) size was used to hold these fasteners together. In the experiments, ambient conditions (relative humidity and temperature) were also measured.

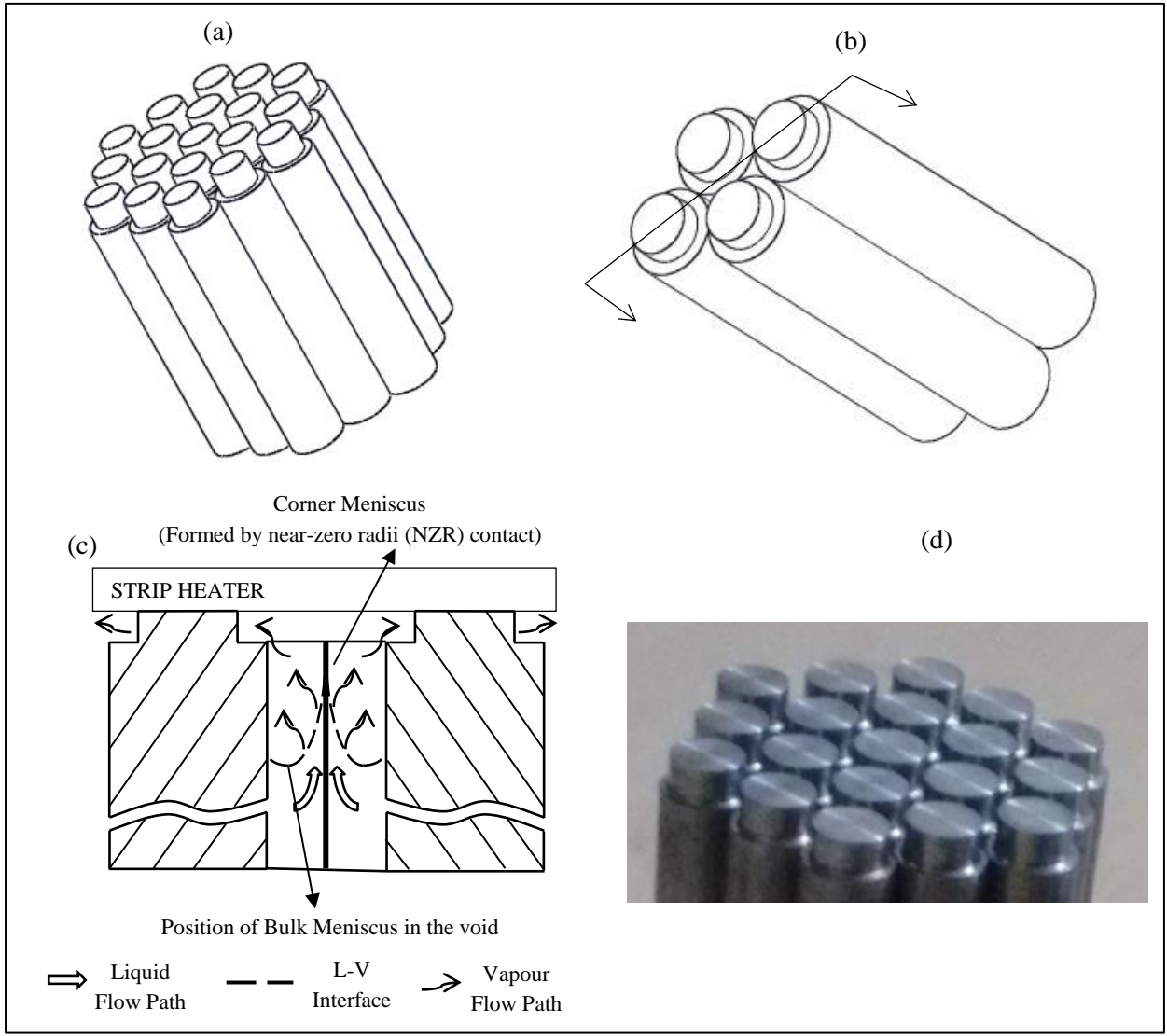

Figure 8 (a) Wick arrangement, (b) a unit cell consisting of four rods, (c) cross-sectional view, and (d) three layers of rods i.e. 19 rods in total with grooves near their top. The wick thus consists of packed smooth stainless steel rods machined at one end so as to provide an escape route for the evaporated vapours. *(a-c) are taken from [30].

Another important aspect of the conductive study is in designing the vapour escape routes. The heater is pasted to the top surface of the rods and we need to provide some space here for the vapours (the side and bottom surfaces are impervious). For this purpose, the top ends of the rods were machined (see Figure 8a). The final structure of the rods is: 4 mm diameter of (SS) pillars having 2 mm depth on top of 73 mm length SS rods with 5 mm diameter (the top of the rods are seen in Figure 8d). Figure 8b shows a unit cell consisting of four machined SS rods. A cross-sectional view along the direction marked in Figure 8b is seen in Figure 8c. Strip heater heats the top surface of these pillars first. Most of the heat then gets conducted, via TIM, downwards (2 mm) to the LV meniscus. The liquid would now rise to a height of 73 mm, where we have a sudden change in the diameter; this was confirmed separately when two rods were dipped in a pool of coloured water. This configuration ensures smooth release of the evaporated vapours. Routes of different phases i.e. the liquid and vapours pathways are shown in Figure 8c. In order to minimize the heat loss to the ambient, glass wool, wrapped around the wick, was used as an insulation material as seen in Figure 7. In the final set of experiments, the brass tank was also insulated at the sides. The insulation diameter was ~8.5 cm and height ~9 cm. The whole process mimics the real environment of an LHP except the condenser region, which is unavailable in the present experiments. The source tank can be thought of a scaled version of the one used in LHP.

## 4. Results and discussion

In this section, first, we briefly discuss the scales of corner film velocity in the NZR of contacts obtained from a separate experiment. We then discuss the evaporating or cooling characteristics, at moderately high heat loads, of the proposed wick consisting the nineteen packed rods. Heat budget if the process is then discussed; apart from some heat lost in changing the phase of the liquid, where is the rest heat going? We close this section by showing some results from the numerical simulation.

*Liquid film velocity in 'the rate of the liquid rise experiment'*

Two 3 mm diameter glass rods (75 mm length), touching each other, were kept vertical such that the gap '$d$' $\to 0$. The rods' bottom was dipped in a pool of (coloured) water and the rate of liquid rise was filmed; the water level was varied in this experiment to get better contrast. Liquid rising in the NZR of contact was seen as a thin vertical black line. We calculated the corner film velocity $(v_f)$ of ~10 cm/s at $Z = 30$ mm. Eq. (10) gives $v_f = 22$ cm/s at $Z = 30$ mm. These values are of the same order obtained in the experiments. Some difference could be due to the reduced surface tension (due to $KMnO_4$ particles) in the experiment. This high rate of liquid velocity in the proposed wick is unlikely in the conventional wicks.

*Results from evaporation experiments*

Experiments were conducted for two different evaporating liquids: (1) deionized water and (2) highly volatile n-pentane. The liquid is poured to about 25 mm height in the source tank, while the valve stops its flow to the brass tank (initially it does not have liquid). This corresponds to about 250g of water and 157g of n-pentane. After the valve is opened the liquid level in both the tanks become equal (~17 mm). In both the cases viz. with water and n-pentane, the bulk meniscus was kept ~55 mm far from the rods' top end. The experiment is commenced when the weighing scale showed a stable value such that any force exerted by the strip heater did not enter our final observations. Pressure above the liquid level in the source tank was maintained atmospheric via a breather hole on its top surface. We first discuss the mass loss curves from different experiments and then will discuss the wick characteristics in the steady state.

The total mass loss, including from the source tank, at different variac voltages, are seen plotted versus time for the cases of water in Figure 9. Experiment with water was commenced at 110V ('circles' in Figure 9). In the beginning, the mass lost increases slowly (heating increases the operating temperature gradually, which in turn increases the rate of evaporation) with time but eventually varies linearly with time in the steady state, within 1h, signifying a constant evaporation rate. The variac voltage was changed in the order - 110V → 165V → 55V → 80V → 130V → 30V; this was done to study the effect of abrupt changes in the incident heat load. Data for 165V is not shown since at this voltage the TIM started burning. The initial transience in all the curves is due to the sudden change in the input heat load prior to it. Table 1 summarizes the experimental study with water. Drop in the liquid level (fourth column in Table 1), in the source tank, was estimated using the total mass loss and the continuity equation. The liquid drop in all the cases is small compared to the initial level (~17 mm) in the source tank. This small liquid level drop should not have affected the corner film geometry. A significant drop in the liquid level would lead to higher distance between the bulk meniscus and the rods' top, which would change the corner velocity scales. However, it was previously observed [18] that even if the liquid level in the voids of the wick goes down by 70 mm, the evaporation rate is still about 90% of the initial evaporation rate (~400 mm/day) when the wick was completely wet.

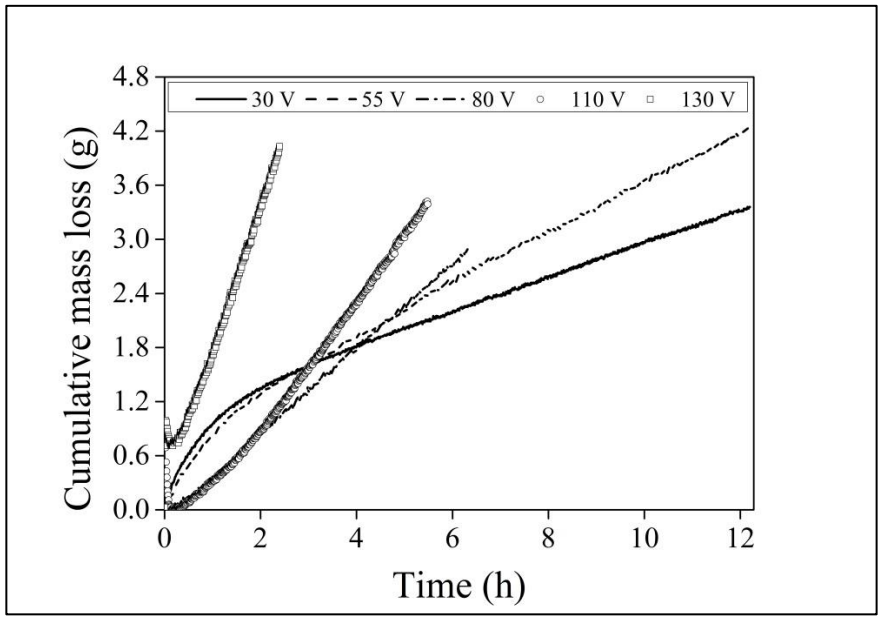

Figure 9 Variations, versus time, of the experimentally measured cumulative mass losses at different variac voltages or heat inputs. The evaporating liquid is deionized water.

Table 1 Experimental data with water as the evaporating liquid.

| Run | Voltage (V) | Duration (h) | Mass loss (g) | Liquid level drop (mm) |
|---|---|---|---|---|
| 1 | 110 | 5.48 | 3.39 | 0.29 |
| 2 | 55 | 12.22 | 4.22 | 0.36 |
| 3 | 80 | 6.32 | 2.89 | 0.25 |
| 4 | 130 | 2.4 | 4.03 | 0.35 |
| 5 | 30 | 12.2 | 3.36 | 0.29 |
| 6 | 0 | 13.54 | 0.54 | 0.05 |

Similarly, experiments were conducted with n-pentane as the evaporating liquid; Figure 10 shows the variations, with time, of different cumulative mass loss curves. The variac voltage was changed in the order - 0V → 30V → 55V → 80V → 110V → 130V. The initial liquid level in all the cases here was kept the same as that used in the experiments with water. Unlike the water case, steady state is achieved much faster in this case, within 20 minutes. An experiment was performed with the heater switched off. The required heat for latent heat release comes from the hotter ambient. We observed only a small difference between the curves for cases with 0V and 30V represented by solid and dashed lines, respectively, in Figure 10. This suggests that at low heat loads the wicks' behaviour is similar to that when no external heating is applied. Further, unlike the water case, the drop in the liquid level with pentane (fourth column in Table 2) is considerably higher and thus extra pentane was periodically poured in the source tank to ensure the fixed initial liquid level (55 mm from the rods' top).

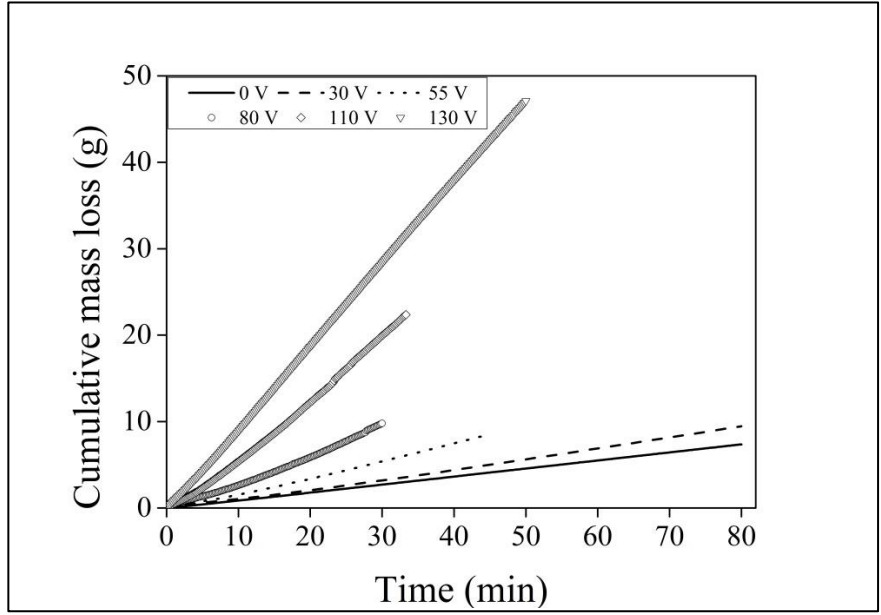

Figure 10 Variations, versus time, of the experimentally measured cumulative mass losses at different variac voltages or heat inputs. The evaporating liquid is n-pentane.

Table 2 Experimental data with n-pentane as the evaporating liquid.

| Run | Voltage (V) | Duration (min) | Mass loss (g) | Liquid level drop (mm) |
|---|---|---|---|---|
| 1 | 0 | 80 | 7.35 | 1.01 |
| 2 | 30 | 80 | 9.46 | 1.30 |
| 3 | 55 | 45 | 8.54 | 1.17 |
| 4 | 80 | 30 | 9.79 | 1.35 |
| 5 | 110 | 33.33 | 22.38 | 3.09 |
| 6 | 130 | 50 | 47.10 | 6.49 |

In the steady state (after about 60 and 20 min with water and n-pentane, respectively), the mass loss is linear and thus the evaporation rate is a constant. Note that mass loss happens from the wick and the source tank both. A separate experiment was conducted, where only the source tank was allowed to evaporate, while the liquid (n-pentane) level was maintained the same as in the previous experiments. This experiment may not have operated at the same temperature as in the actual experiments but for the simplicity, we have neglected heat transfer due to siphoning here. The mass loss from the wick was then calculated by subtracting the source tank mass loss data from the total mass loss. Evaporation rate was obtained by taking the area of the regular hexagon whose vertices represent the centres of the outermost 5 mm diameter rods; this area is ~3.97 cm$^2$. The obtained steady state evaporation rates are plotted versus the heat load in Figure

11. The maximum achievable evaporation rate with water (filled marker) was ~73 mm/day (~1.25 cm$^3$/h) at 2.79W/cm$^2$ heat load (or 130V) but at the same voltage, with pentane, the evaporation rate was ~75 times (5500 mm/day or ~94.12 cm$^3$/h) higher. Without external heating, the evaporation rates were about 3 mm/day and 370 mm/day for water and pentane, respectively.

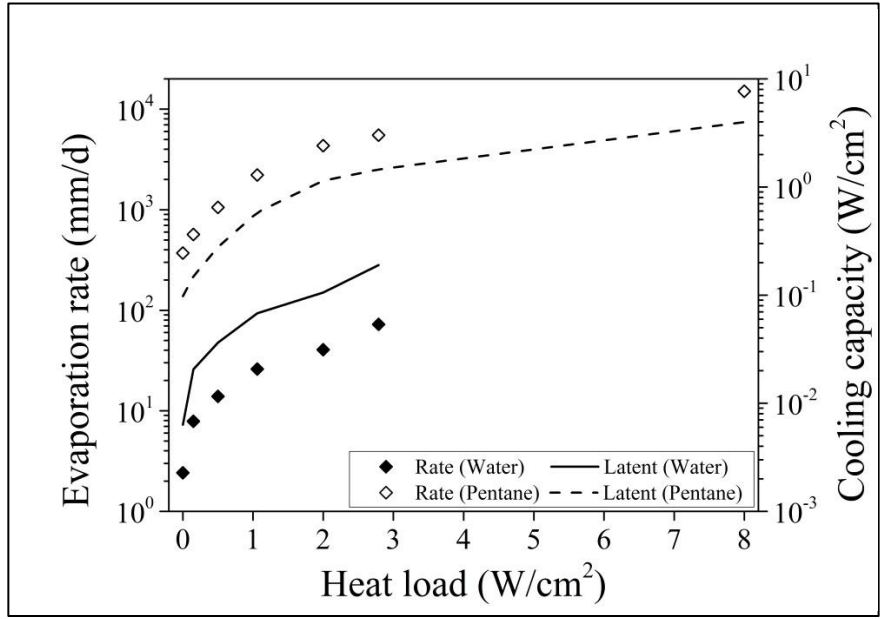

Figure 11 Steady state evaporation rates from the wick with water and n-pentane. Symbols and lines represent the rate of evaporation and the latent heat loss (cooling capacity), respectively. Logarithmic scales have been used for both the vertical axes.

Table 3 Cooling capacity of the wick with two different liquids at different expected heat loads.

| Voltage (V) | Heat Load (W/cm$^2$) | Cooling capacity (W/cm$^2$) Water | Cooling capacity (W/cm$^2$) n-Pentane |
|---|---|---|---|
| 0 | -- | 0.006 | 0.098 |
| 30 | 0.15 | 0.021 | 0.150 |
| 55 | 0.50 | 0.037 | 0.278 |
| 80 | 1.06 | 0.068 | 0.584 |
| 110 | 2.00 | 0.106 | 1.143 |
| 130 | 2.79 | 0.139 | 1.455 |
| ***220*** | ***8.00*** | ***---*** | ***3.976*** |

The evaporated mass (or rate of evaporation) is converted into the latent heat loss (or cooling capacity) using the latent heat of vaporization of the corresponding liquid. The variation of the cooling capacity of the wick is plotted along the secondary vertical axis versus the heat load (Figure 11). With increasing the heat load, the cooling capacity increases almost proportionally; this is the effect of the sustained corner films in the proposed wick. This is a remarkable and unique feature. Rated power of the strip heater was 50W at 220V (or 8W/cm$^2$). The other heat loads were estimated assuming a power law (see Table 3). Wick with n-pentane as the evaporating liquid performs much better compared to that with water. At a heat load of ~2.8W/cm$^2$, the wick was able to release ~1.5W/cm$^2$ in the form of latent heat. The measured groove temperature (just below the TIM and between the rods) in the steady state, in this case, was ~70°C. At full power (8W/cm$^2$), the cooling capacity, with n-pentane, is ~4W/cm$^2$; this experiment was conducted for a short duration; data for this case is not included in Figure 10. It is noteworthy that this cooling capacity is within the maximum permissible cooling performance (5.4W/cm$^2$ at Z=55mm for 5mm diameter rods and pentane as the evaporating liquid; see Table A1 in Appendix A). Note that the top face of the strip heater was uninsulated and the cooling capacity may further improve after insulation. With the existing strip heaters available and the exposed area used in the present experiments, the heat load could not be increased further. It would be interesting to increase the heat load to ~100W/cm$^2$ and test this proposed wick's performance. As per the theoretical data (Appendix A), 1 mm diameter and 10 mm long rods can release ~10$^4$W/cm$^2$ (or 10$^8$W/m$^2$) in latent heat. At these extremely high heat loads, we recommend the wicks' to be shorter and the bulk meniscus nearer to the evaporating end. Further, it would be interesting to observe changes in the evaporation behaviour at these heat loads; boiling may take over as the primary mechanism of phase change and dry-out may occur at this point with the proposed wicks.

A short mention of the physics of the evaporation phenomenon is necessary and we briefly discuss it. The LV meniscus which is pinned at the NZR of contact receives the heat mostly via conduction through the rods. The region of the corner film near the rod can be divided into three zones [31-33] (a) a thin adsorbed (non-evaporating) zone, (b) an evaporating zone, and (c) a near-saturated zone (this acts as a liquid source to the evaporating zone). For a flat wall, it was argued that the evaporating zone would be limited to a short (~10$\mu m$) length scale. However, for a curved surface like a rod, this length scale may not be true. Additionally, it was shown [18] that, in case of packed rods, evaporation contribution may come from deeper regions too. A unique method for modelling evaporation [18] was adopted based on the measured evaporation rate, top, and bottom surface temperatures. Details of a few other models can be found elsewhere [34].

*Heat budget*

We discuss, in some detail, different heat (loss or gain) terms associated with the evaporation process in the (pentane) experiment with 130V variac voltage. A schematic (Figure 12) shows different heat terms along with the experimental components along a central plane orthogonal to the vertical direction. Heat ($Q_{in}$[W]) is received by the top of the rods (pillar portion) by the strip heater. In the steady state, heat balance for the strip heater can be written as

$$Q_{supply} = Q_{in} + \left(Q_{rad}^h + Q_{conv}^h\right) + \left(Q_{rad}^m + Q_{conv}^m\right) \qquad (20)$$

Part of the power supplied by the heater $\left(Q_{supply} = 17.44W\right)$ is lost into the ambient by convection and radiation from the top side of Mica $\left(Q_{rad}^m + Q_{conv}^m = 1.13W\right)$ and heater $\left(Q_{rad}^h + Q_{conv}^h = 2.47W\right)$ (see Figure 12). These values were calculated (see Eqs. C1-C4 in Appendix C) based on the materials' dimensions (Figure 7 and Figure 12a) and the (average) surface temperatures (Figure C1). Standard emissivity values were used for the heater (0.10) and mica (0.75) in these calculations. Radiation and convection terms were estimated following *Stefan-Boltzmann law* and *Nu-Ra* correlation for heated horizontal surfaces [35]. Details of the calculations are seen in Appendix C.

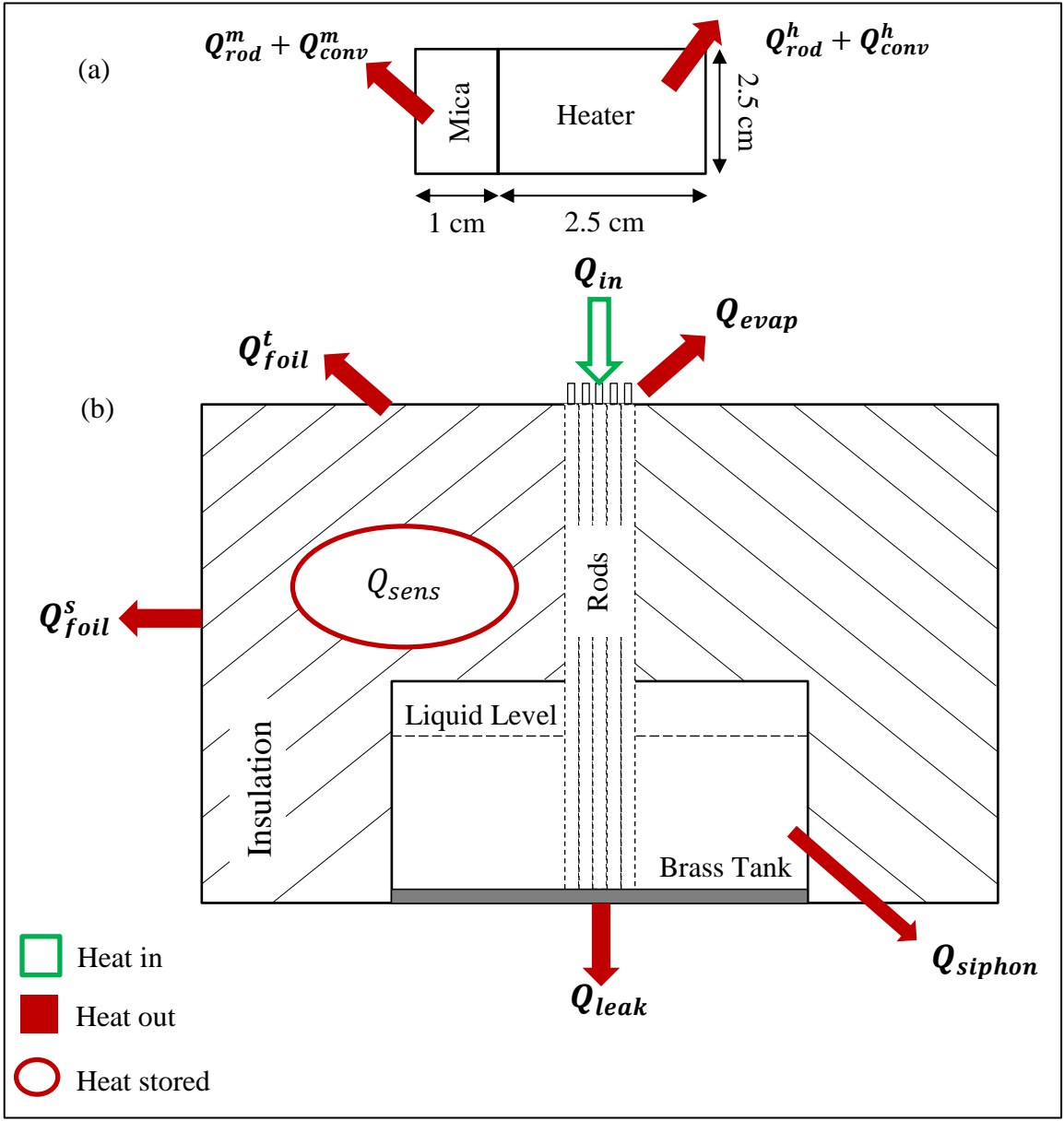

Figure 12 Different heat flux terms associated with the system during evaporation from the experimental setup.

We estimated $Q_{in}$ (=13.84W) following the heat losses and Eq. (20). $Q_{in}$ is the incident heat load on the wicks, which is distributed into different heats as

$$Q_{in} = (Q_{evap}) + (Q_{foil}^t + Q_{foil}^s) + (Q_{leak} + Q_{siphon}) + [Q_{sens}] \quad (21)$$

Where, $Q_{evap}$ is the latent heat loss term, $Q_{insul}^t$ and $Q_{insul}^s$ are the (total) heat losses from the top and side surfaces of the insulation, covered with aluminium foil (emissivity of 0.05), respectively, $Q_{leak}$ is the heat loss from the bottom surface of the brass container, $Q_{siphon}$ is the heat loss from the experimental container to the source tank via the siphon tube, and $Q_{sens}$ is the sensible heat term, which is consumed in raising the temperature of the entire system. Using Eq. (C9) and Eq. (C12), we estimated $Q_{insul}^t$ and $Q_{insul}^s$ to be nearly 1.34 and 0.52W, respectively following [36] correlation for a heated vertical surface (side insulation here). We assume $Q_{leak} \sim Q_{insul}^S = 0.52$W and neglect $Q_{siphon}$, which is supposed to be very small. Further, we calculated $Q_{evap} \cong 5.78$W using Eq. (C11). The detailed calculation for these terms is seen in Appendix C. Using Eq. (21), we obtained $Q_{sens} = 5.68$W. The entire idea for conducting the heat budget was to get the sensible heat term. Using this value, we estimated the average rise in the system temperature to be ~23.5$^0$C (Eq. C18). The initial temperature was ~23.5$^0$C (the ambient temperature, $T_{amb}$ varied between 23$^0$ and 24$^0$C during the experiment) and thus the final average system temperature is expected to be ~47$^0$C, which seems reasonable. In the steady state, we expect improvement in the cooling performance of the wick since the sensible term would be zero.

*Results from simulations*

We performed numerical simulations (COMSOL MULTIPHYSICS) to study the effect of the rod diameter and the material thermal conductivity on the wick performance. One major question is how the resistance to heat transfer is affected by these parameters? The other goal is to compare the temperatures obtained in the experiments with those from the simulations. We solve the 3-D unsteady heat conduction equation in a single rod (Figure 13). This rod is surrounded by 6 rods, as in the experiments, which, in simulations, have been suitably replaced by 6 liquid films at the periphery. Each film subtends an angle of 20$^0$ at the centre; this angle was fixed based on the geometrical conditions in the experiments. The initial temperature in all the simulations was 25$^0$C i.e. the ambient temperature.

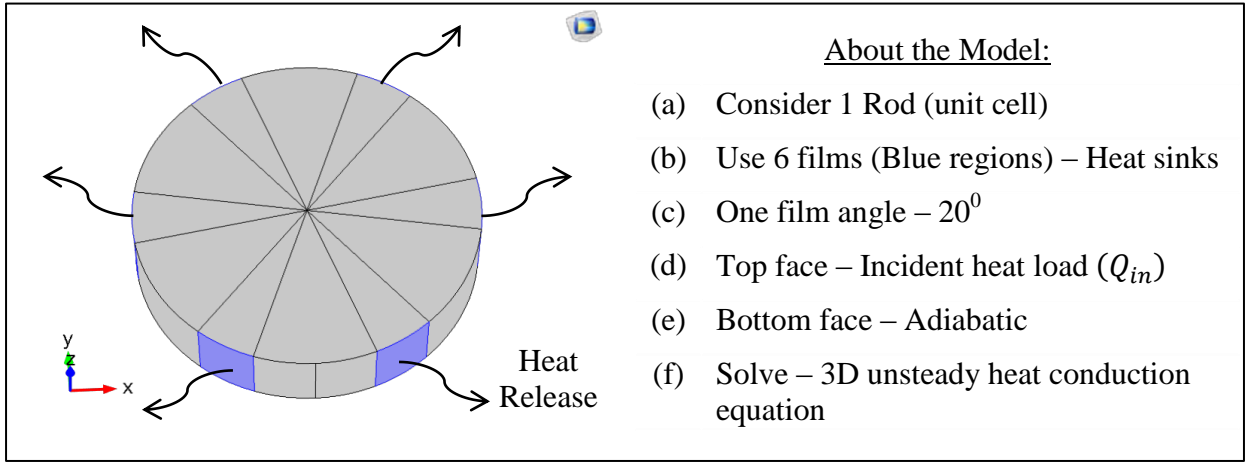

Figure 13 Schematic of the control volume used for the numerical simulation. Blue regions represent the corner films. Boundary conditions are seen on the right side.

Two conditions were used at the rod's top surface: (a) a constant incident heat flux ($Q_{in}$) and (b) a convective heat transfer coefficient (equal to 1W/m$^2$-K) for incorporating interaction with the ambient. In the experiments, this face of the rod(s), is glued through a TIM to the strip heater, interacted with the ambient via the heater. We used a *heat sink condition* (corresponding to the measured latent heat loss in the experiments) at the film locations; this input is very important in the present model. The rod's bottom surface was assumed adiabatic, while symmetry condition was used at the remaining rods' regions (non-film zone). In the simulations, we only considered the near-top region for computing the temperature distribution since, previously [9,18], we have observed that most of the evaporation occurs near the exposed end.

The film length considered for the simulations was 1 mm, which is much higher than the evaporating film length obtained in case of the flat plates [31].

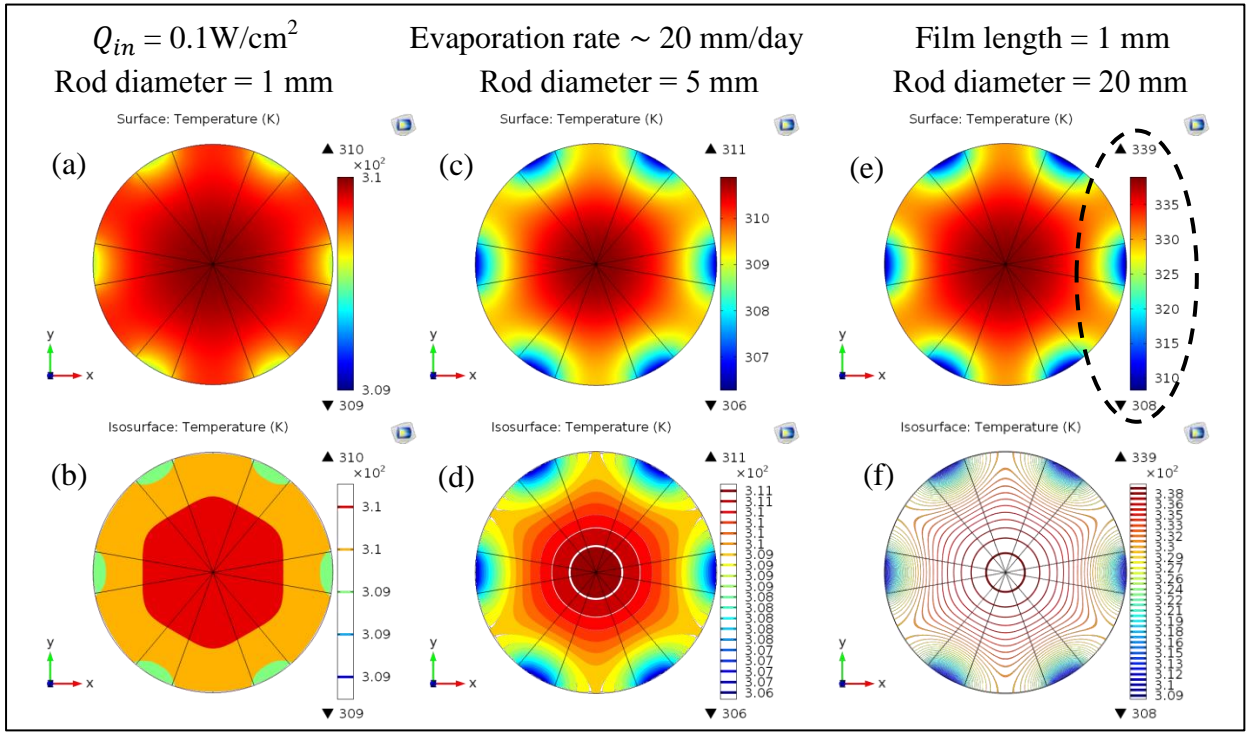

Figure 14 Distribution of the top surface temperatures with the temperature isosurfaces for three different glass rod diameters. Conditions used in the simulations are mentioned at the top of the figure. The evaporating liquid is water.

First, we discuss results on the effect of rod diameter in case of a low heat load (~0.1W/cm$^2$). Figure 14 shows the distribution of the top surface temperature and the temperature isosurfaces for three different (glass) rod diameters 1, 5, and 20 mm. As expected the hottest point is the rods' centre, while the coldest point is the film regions since these are the sources of heat release. The difference in the top surface temperature between the centre and the edge increases with rod diameter (~1$^0$C for 1 mm diameter, ~5$^0$C for 5 mm diameter, and ~30$^0$C for 20 mm diameter case). Clearly smaller diameter rod will have lower temperature at the heat source, which is desirable.

We now discuss the effect of rods' material on the temperature distribution. For this case we chose 5 mm diameter rods' diameter, while the materials chosen were stainless steel and glass; these simulations were performed for ~2.79W/cm$^2$ heat load case corresponding to $Q_{in} = 13.84W$. Note that the thermal conductivity of stainless steel is an order of magnitude higher than that of glass. Figure 15 shows the top surface temperature distribution along with the temperature isosurfaces. Two major differences can be clearly seen. First, for the stainless steel case, the top surface temperature is nearly uniform with a small temperature gradient (Figure 15b) from the centre to the edge. The top surface temperature in the stainless steel case is ~15$^0$C lower than the glass case. Note that for the same conditions, we measured the groove temperature (using a T-type thermocouple placed just below the TIM and in between the rods' top region) to be ~70$^0$C. The bottom surface temperature, for the stainless steel case, in the numerical simulation was ~66$^0$C, which is very close to the measured temperature value. Second, due to the high cooling requirement, the coldest temperature in the glass case is ~21$^0$C, which is

~$46^0$C lesser than the stainless steel case. In fact, this temperature value is lesser than the initial value (ambient temperature) of $25^0$C used in the simulations. For 5 mm diameter rods, a temperature drop of $62^0$C in the top surface from the centre to the film seems highly unlike in real situations.

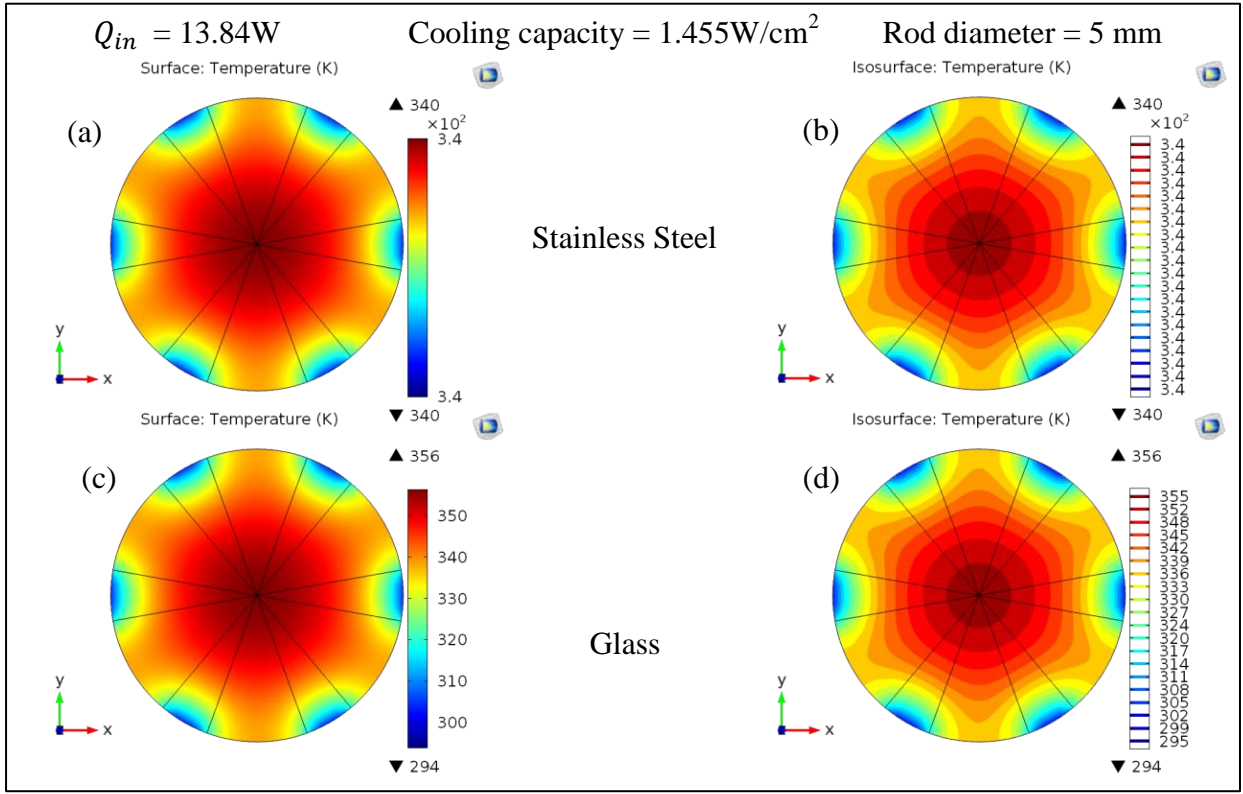

Figure 15 Distribution of the top surface temperatures with the temperature isosurfaces for two different rod materials but the same diameter. Conditions used in the simulations are the same as that in the experiment with pentane (see Table 3).

We conclude from the numerical simulations that it is better to use rods having a lower diameter and higher thermal conductivity. Larger diameter rods offer high resistance to heat conduction and would be a bad choice for the wick. Similarly, a material with a lower thermal conductivity offer more resistance to heat, that is desired, to be conducted quickly.

## 5. Conclusions and possible future

A new type of wick consisting of closely packed circular rods is proposed. The wick was tested, for its cooling performance, between the heat loads of 0.05W/cm$^2$ to 8W/cm$^2$. With water as the evaporating liquid, Silicone-based thermal interface material (TIM) burnt at the higher heat loads; water was found to have bad cooling characteristics. The cooling capacity of the wick with n-pentane was found to be ~10 times better than with water. With pentane, heat loss in the form of the latent heat was ~50% of the applied heat load; heat transfer aspects have been discussed in detail, in Appendix C. At the highest heat load studied experimentally here, i.e. 8W/cm$^2$, the highest temperature reached was ~$70^0$C, this value was also validated in the simulations.

These characteristics of this 'new' wick seems to be better than the conventional spheres-based wicks whose cooling performance depends greatly on the applied heat load. Note that high cooling can only be achieved if the evaporation (or phase change) occurs close to the heat source. The entire idea is, therefore, to retain the LV meniscus near the hot spot at high heat loads as well. The present wick has near-zero radii (NZR) of contacts that are formed between the smooth rods; these contacts are sustained along the rods' length as well. NZR of contacts theoretically lead to infinite capillary rise (Table B1) and hence, during evaporation, the liquid gets pinned along these locations, while the bulk (liquid residing in the space formed between three rods in mutual contact) recedes within the void [18]. Additionally, the liquid rises rapidly (Table A1 and A2) along the NZR of contacts; this ensures fast replenishment of the liquid to the evaporating end. The proposed wicks thus promise even higher cooling performance at very high incident heat loads, where in the extreme cases the conventional wicks may suffer '*dry-out*' i.e. no liquid present in the wicks. The effect of the incident heat load on the evaporation characteristics of a homogenous porous medium has been recently [29] reported.

The new wick, which seems to be better than the regular spheres-based wicks, has many unique features and could be a suitable replacement for them in HP/LHP/CPL. The conceptualization for the possible use of these newly proposed wicks in real situations, viz., in the cooling devices, have been previously [30] explored. A guide for selecting wicks based on the rods' diameter, rods' height, the liquid used, and maximum permissible evaporation demand has also been given (see the Appendices). In future, the wick characteristics need to be tested at much higher heat loads (50-500 W/cm$^2$) for future cooling solutions.

In the present study, the liquid rises against gravity since the rods are packed vertically. However, the rods can also be packed horizontally with their cross-sections exposed; higher rates of evaporation were observed [37] in this orientation from a single rectangular capillary. Drastically reduced rates of evaporation were reported [19] in cases where the lateral side of the rods was exposed to the ambient. Management and effect of liquid volume has also been studied recently [38], where a similar geometry, bundle of closely packed capillaries, was explored.

**Acknowledgements**

We thank Indian Space Research Organization (ISRO) for funding the research under the grant ISTC/MME/JHA/0374. Funding from Ministry of Earth Sciences, India under the grant MESO/0034 is also gratefully acknowledged.

**Appendix A. Film velocities $(v_f)$ and the maximum permissible evaporation rate $(E_l)$**

Since the corner film properties are known along with the corner film velocity, we would be able to estimate the maximum permissible evaporation rate $(E_l)$ using Eq. (14) or Eq. (19) for different $R_{rod}$ and at different $Z$ values. For demonstration, we again consider two values of rod diameters only, 1 and 5 mm. Note that $t, \alpha_Z, v_f,$ and $E_l$ are calculated using Eq. (10), Eq. (6), Eq. (13), and Eq. (19), respectively. Note that $E_l$ is estimated (Figure 5) for two layers of rods. For calculating the cooling capacity $(Q_{evap})$, we use

$$Q_{evap}[\text{W/cm}^2] = (\rho_l \lambda E_l)/10 \quad (A1)$$

Where, $E_l$ is in 'm/s', $\rho_l$ is in 'kg/m$^3$', and $\lambda$[kJ/kg] is the latent heat of vaporization of pentane.

Table A1 shows the values of different parameters for 5 mm diameter rod and pentane as the evaporating liquid. Note that the contact angle has been assumed to be zero degrees In relation to our experiment, the corner film reached till $Z = 73$ mm while the free liquid level at the beginning of the experiment was at $Z \sim 18$ mm. The distance between the rods' top (holding the corner film) and the free liquid level is therefore 55 mm. We must, therefore, look for the values corresponding to $Z = 55$ mm in Table A1. At this $Z$ value, $\alpha_Z \sim 11°$ and the permissible cooling capacity is about 5.4W/cm². Referring back to Table 3, the wick was able to release ~4W/cm² in our experiments, which is within the permissible range. The cooling capacity reduces drastically for longer wicks. For example – 20 mm long rods can sustain a cooling load ~176W/cm², which is about 11 times higher than the rods with twice the length. The corner film velocity increases drastically for shorter rods, it is ~60 cm/s for 10 mm long rods. This rod length can, theoretically, sustain ~1800W/cm² of cooling load; this range of cooling may not be needed in the existing devices. In relation to the cooling devices, rods length may vary from 10 to 50 mm for application to high heat loads.

Table A1. Estimated theoretical permissible evaporation rates and the corresponding cooling capacity at different $Z$ values of rods. $R_{rod} = 2.5$ mm and the evaporating liquid is n-pentane. Experimental conditions are similar to the values mentioned in the italicised row.

| $Z$ (mm) | Time taken, $t$ (s) | $\alpha_Z$ at rods' top (°) | $v_f$ at rods' top (cm/s) | $E_l$ at rods' top (cm/s) | $Q_{evap}$ (W/cm²) |
|---|---|---|---|---|---|
| 10 | 5.76 x 10⁻³ | 25.1 | 57.9 | 8.03 | 1828.2 |
| 20 | 4.61 x 10⁻² | 18.1 | 14.5 | 0.77 | 175.8 |
| 30 | 0.16 | 14.9 | 6.4 | 0.19 | 43.8 |
| 40 | 0.37 | 12.9 | 3.6 | 0.07 | 16.2 |
| 50 | 0.72 | 11.6 | 2.3 | 3.29 x 10⁻² | 7.5 |
| *55* | *0.96* | *11.1* | *1.9* | *2.37 x 10⁻²* | *5.4* |
| 60 | 1.24 | 10.6 | 1.6 | 1.75 x 10⁻² | 4.0 |
| 70 | 1.98 | 9.8 | 1.2 | 1.03 x 10⁻² | 2.3 |
| 75 | 2.43 | 9.5 | 1.0 | 8.06 x 10⁻³ | 1.8 |

Table A2. Estimated theoretical permissible evaporation rates and the corresponding cooling capacity at different $Z$ values of rods. $R_{rod} = 0.5$ mm and the evaporating liquid is n-pentane.

| $Z$ (mm) | Time taken, $t$ (s) | $\alpha_Z$ at rods' top (°) | $v_f$ at rods' top (cm/s) | $E_l$ at rods' top (cm/s) | $Q_{evap}$ (W/cm²) |
|---|---|---|---|---|---|
| 10 | 5.76 x 10⁻³ | 48.9 | 57.9 | 51.77 | 11790.7 |
| 20 | 4.61 x 10⁻² | 37.5 | 14.5 | 6.31 | 1436.9 |
| 30 | 0.16 | 31.6 | 6.4 | 1.73 | 393.2 |
| 40 | 0.37 | 27.8 | 3.6 | 0.67 | 153.4 |
| 50 | 0.72 | 25.1 | 2.3 | 0.32 | 73.1 |
| 55 | 0.96 | 24.0 | 1.9 | 0.23 | 53.2 |
| 60 | 1.24 | 23.1 | 1.6 | 0.17 | 39.7 |
| 70 | 1.98 | 21.5 | 1.2 | 0.10 | 23.6 |
| 75 | 2.43 | 20.8 | 1.0 | 0.08 | 18.7 |

Similarly, the values of different parameters are seen in Table A2 for 1 mm diameter rods. It seems that the permissible cooling capacity, in this case, increases 10 fold compared to 5 mm diameter rods. Note that the corner film velocity is the same in both cases. The enhanced capacity is a direct result of increased $\alpha_Z$ which increases the area of the film available for evaporation. Reducing the diameter of rods further ($R_{rod} < 0.5$ mm) might be difficult from a viewpoint of fabrication. Note that a similar table can be obtained for different evaporating liquids, non-zero contact angles, and various rod diameters. Table A1 and A2 should not be used if the rods are not in physical contact. These values will remain absolutely the same in cases with various numbers of layers of rods given the structure is closely packed with the least porosity. Note that at higher liquid film velocities, the viscous forces might not be negligible and hence must be included in the formulation for the required pressure drop.

**Appendix B. Impact of rods not in contact**

It is very important, in the proposed wick, that the rods are in contact with one another to form NZR of contacts everywhere. If the rods are separated by some finite distance (Figure B1), the height till which the liquid rises would be limited. In worst cases, this rise height might be smaller than the rod lengths. This situation is similar to the one depicted in Figure 2, where the evaporating meniscus is away from the heating source. The wick cannot sustain high heat loads in such cases. Table B1 shows some of these features.

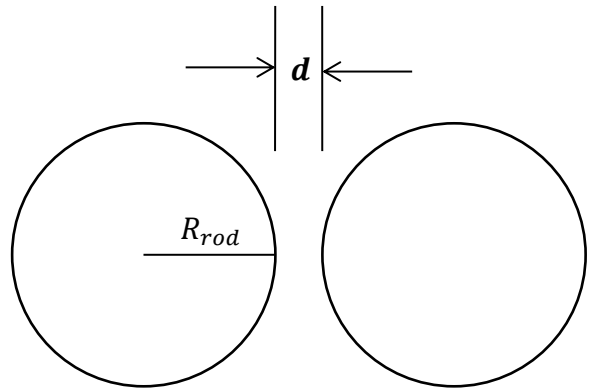

Figure B1. A schematic showing two rods of radii $R_{rod}$ separated by a finite distance $'d'$.

Table B1. Values showing the effect of a finite gap between the rods. Small gaps may represent the roughness and flatness levels of the rods. These calculations are for pentane as the evaporating liquid.

| $d/R_{rod}$ | $\alpha_Z$ (°) | $R_{rod}/r_Z$ | $R_{rod}$ / h | $R_{rod}$ / h |
|---|---|---|---|---|
| → 0 | → 0 | ∞ | 0.5 mm / ∞ | 2.5 mm / ∞ |
| 0.002 | 0.81 | 910.2 | 0.5 mm / 4.74 m | 2.5 mm / 0.95 m |
| 0.02 | 4.09 | 79.50 | 0.5 mm / 0.41 m | 2.5 mm / 8.29 cm |
| 0.2 | 23.03 | 5.121 | 0.5 mm / 26.68 cm | 2.5 mm / 5.33 mm |
| 0.4 | 39.92 | 1.771 | 0.5 mm / 9.23 mm | 2.5 mm / 1.85 mm |
| 1.0 | 81.64 | 0.1073 | 0.5 mm / 0.56 mm | 2.5 mm / 0.11 mm |

As seen in Table B1, the separated corner has a drastic effect on the maximum height rise of the liquid film. When $d \sim 0.2\%$ of $R_{rod}$, pentane would rise to about 5 and 1 m for 1 and 5 mm diameter rods, respectively. For $d \sim R_{rod}$ cases, there is hardly any liquid rise since $Z \sim 0.1$ mm only. Note that in the case of finite separation, the rate of liquid rise will not follow Eq. (10) since this equation is strictly for a pure corner. Now that the height rise would depend on both $d$ and $R_{rod}$, the wick is expected to have properties similar to a conventional one. In the present experiments, $R_{rod} = 2.5$ mm and rods' mean roughness can be assumed to be $O(10\mu m)$ such that even in close contact we may have $d/R_{rod} \sim 0.01$ and the permissible rise of pentane ~20 cm; this is much higher than the rods' length of 7.5 cm. For better performance, smooth, flat, and lower diameter rods are therefore recommended. Note that the system can also be extended to using different sized rods in a wick, where the number of NZR of contacts might not exist everywhere but their numbers would be substantially higher. For similar reasons, rods and plate assembly may also be used as a wick; one such study [39] where only the plates were closely placed has been reported recently.

## Appendix C. Detailed description of the individual terms in the heat budget

We give a detailed description of the heat budget mentioned earlier. We first mention the dimensions of the different materials. Strip heater used was 2.5 cm x 2.5 cm, (exposed) mica was 2.5 cm x 1 cm, insulation at the top surface had a diameter of ~8.5 cm, and insulation on the side was ~9 cm long. The top average surface temperature of the heater, $T_h$ was ~490K, the top average surface temperature of mica, $T_m$ was ~449K, and top average surface temperature of the aluminium foil, $T_a^t$ was ~333K, see Figure C1 for these values. The emissivities were 0.10, 0.75, and 0.05 in the same order. We use these values to estimate heat losses from the top surfaces as

$$Q_{rad}^h = \sigma \epsilon_h A_h (T_h^4 - T_{amb}^4) \tag{C1}$$

$$Q_{rad}^m = \sigma \epsilon_m A_m (T_m^4 - T_{amb}^4) \tag{C2}$$

$$Q_{conv}^h = h_h A_h (T_h - T_{amb}) \tag{C3}$$

$$Q_{conv}^m = h_m A_m (T_m - T_{amb}) \tag{C4}$$

where $\sigma$ is the *Stefan-Boltzmann* constant, $\epsilon_h$ and $\epsilon_m$ are the emissivities of the heater and mica surfaces, $A_h$ and $A_m$ are the exposed areas of the heater and mica surfaces, $T_{amb}$ is the ambient temperature, $h_h$ and $h_m$ are the convective heat transfer coefficients for the heater and mica surfaces. The convective heat transfer coefficients were estimated using

$$h = (Nu)(L^*/k_{air}) \text{ with } Nu = \gamma(Ra)^\delta \tag{C5}$$

where $Nu$ is the Nusselt number, $L^*$ is the characteristic length defined as the ratio of the exposed area to the perimeter of the horizontal heated surface, and $\gamma$ and $\delta$ are the pre-factor and the exponent of *Nu-Ra* correlation. Rayleigh number $(Ra)$ is defined as

$$Ra = \frac{g\beta(T - T_{amb})L^{*3}}{\nu\alpha} \tag{C6}$$

where $g$ is the gravitational acceleration of the earth, $\beta$ is the volumetric expansion coefficient of air, $T$ is the temperature of the concerned surface, $\nu$ and $\alpha$ are the kinematic and thermal diffusivities of air. Following *Nu-Ra* correlations were used

$$Nu = 0.59Ra^{0.25}; 2.0E02 \leq Ra \leq 1.0E04 \tag{C7}$$

$$= 0.54Ra^{0.25}; 2.2E04 \leq Ra \leq 8.0E06 \tag{C8}$$

The total heat $(Q_{foil}^t)$ loss from the aluminium foil (over the insulation) at the top is calculated as

$$Q_{foil}^t = A_f^t \left[\sigma\epsilon_f \left(T_f^{t^4} - T_{amb}^4\right) + h_f^t(T_f^t - T_{amb})\right] \tag{C9}$$

where $A_f^t$ is the exposed surface area, $\epsilon_f$ is the emissivity, $T_f^t$ is the average surface temperature, and $h_f^t$ is the convective heat transfer coefficient of the aluminium foil at the top. Eqs. (C5-C8) were used to estimate $h_f^t$, while $A_f^t$ is calculated as

$$A_f^t = \pi r_{in}^2 - (A_h + A_m) \tag{C10}$$

where $r_{in}$ is the radius of the insulation, equal to ~42.5 mm here. The latent heat term, $Q_{evap}$ is calculated as

$$Q_{evap} = (dm/dt)\lambda \tag{C11}$$

where $dm/dt$ (g/s) is the rate of mass loss and $\lambda$ (J/g) is the latent heat of vaporization of the evaporating liquid.

At last, we give the method for estimating $Q^s_{foil}$ i.e. the total heat loss from the aluminium foil, at the side, wrapped around the side insulation enveloping the rods and brass container. This term can be written as

$$Q^s_{foil} = A^s_f[\sigma \epsilon_f (T^{s\,4}_f - T^4_{amb}) + h^s_f (T^s_f - T_{amb})] \tag{C12}$$

where $A^s_f$ is the exposed surface area, $T^s_f$ is the height-wise average surface temperature, and $h^s_f$ is the convective heat transfer coefficient of the vertical aluminium foil at the side. The exposed area can be calculated as

$$A^s_f = 2\pi r_{in} H_{in} \tag{C13}$$

where $H_{in}$ is the height of the insulation layer, equal to 9 cm here. Note that $T^s_f$ was not measured and it needs to be estimated. We equate the heat transferred from the inner central region of the system to the insulation outer layer to $Q^s_{foil}$, which can also be written as

$$Q^s_{foil} = k_f (T_c - T^s_f)/t_{in} \tag{C14}$$

where $k_f$ is the thermal conductivity of glass wool, $T_c$ is the height-wise averaged temperature of the outermost layer of the rods, and $t_{in}$ is the insulation thickness. In order to estimate $T_c$, we need the height-wise temperature profile, which is nearly impossible to measure in the present experimental setup. We thus assume $T_c$ to be the average of the measured groove temperature, $T_g$ and the bottom surface temperature ($T_{bot}$) of the brass tank. We observed that between t = 45 min to t = 50 min, $T_g \sim 70°C$ and $T_{bot} \sim 30°C$ thus giving $T_c \sim 50°C$. The other unknown in Eq. (C12) is $h^s_f$, which for a heated vertical surface can be obtained using [36]

$$\overline{Nu_H} = 0.68 + \frac{0.67\, Ra_H^{\frac{1}{4}}}{\left[\left\{1 + \left(\frac{0.492}{Pr}\right)^{\frac{9}{16}}\right\}^{\frac{4}{9}}\right]}; \quad Ra_H < 10^9 \tag{C15}$$

where $\overline{Nu_H}$ is the height-averaged Nusselt number and $Pr$ is the Prandtl number. The Rayleigh number, $Ra_H$ is calculated as

$$Ra_H = \frac{g\beta_s (T^s_f - T_{amb}) H^3_{in}}{\nu \alpha} \tag{C16}$$

where $\beta_s$ is calculated at the mean of $T^s_f$ and $T_{amb}$. Eqs. (C12) and (C14) was solved iteratively to get $T^s_f$, we calculated it to be equal to ~302K, which in turn gives $Q^s_{foil} \sim 0.52$W. The last term in the heat budget is the sensible heat term $Q_{sens}$, which may be written as

$$Q_{sens} = \left(m_b C_{p_b} \Delta T_b\right) + \left(m_r C_{p_r} \Delta T_r\right) + \left(m_l C_{p_l} \Delta T_l\right) + \left(m_{in} C_{p_{in}} \Delta T_{in}\right) \tag{C17}$$

where $m$, $C_p$, and $\Delta T$ represent mass, specific heat capacity, and temperature increment, respectively. The subscripts $b, r, l$, and $in$ represent brass tank, rods, evaporating liquid, and insulation respectively. Eq. (C17) can be simplified, assuming similar order of temperature increments, as

$$Q_{sens} = \left(m_b C_{p_b} + m_r C_{p_r} + m_l C_{p_l} = m_{in} C_{p_{in}}\right)(\Delta T_{sys})/t \tag{C18}$$

where $\Delta T_{sys}$ is the increase in the overall system temperature and $t$ is the elapsed time. Using Eq. (21) and calculated values of other heat terms we estimated $\Delta T_s \sim 22.5^0 C$ through Eq. (C18). Note that Eq. (C17) or (C18) neglects the thermal masses of the heater, Hylam fasteners, and the insulation; these are negligible compared to the other important contributors. Mass of brass tank, stainless steel rods, insulation, and liquid pentane used during the experiment were 995.30, 215.45, 167.15, and 45.77 g, respectively. The materials' specific heat capacities in the same order are 0.38, 0.50, 0.84, and 2.32 J/g-K. The sum of all the thermal masses was 732.6 J/K.

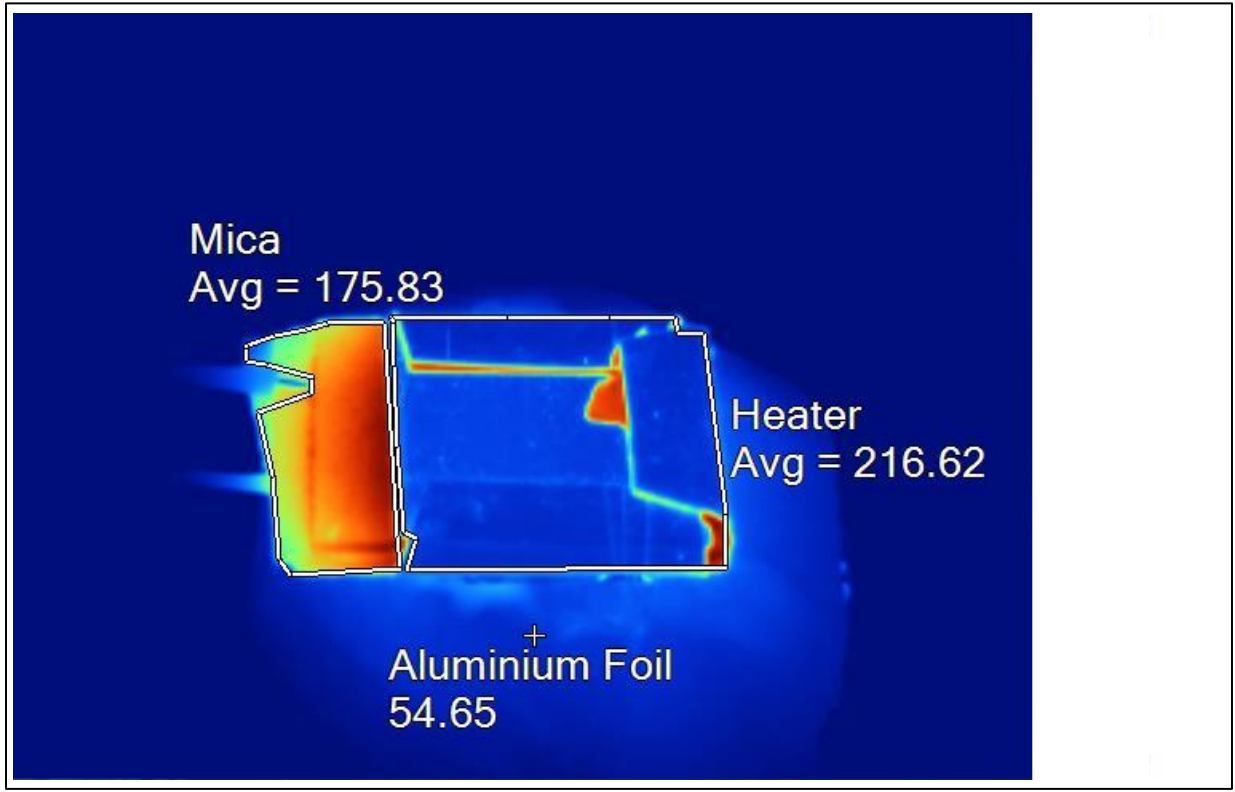

Figure C1. A thermal image showing the top surface temperature distribution of different components. For heater and mica, the average temperatures are seen. The marker represents the temperature of the aluminium foil, which was above the insulation.

Figure C1 shows the temperature distribution of the top surface captured using a thermal camera. Note that in a radiometric image, captured by a thermal camera, the emissivity value is fixed; Figure C1 was captured when emissivity was set to 0.80. There are however three different materials (a stainless steel strip heater, a bare mica piece, and aluminium foil) in Figure

C1; all with different emissivities. During the post-processing, we changed the emissivities for the respective materials with their exact values.

**Appendix D. Liquid flow velocity scale in steady state and viscous pressure loss**

As mentioned previously, two velocity scales exist, in the present investigation, each corresponding to a transient phase (during capillary rise) and a steady state phase (when the bulk meniscus gets a stable position for a particular heat load). In the steady state, a phase change occurs at the evaporator end while the corner films keep supplying the required amount of liquid to this end. The liquid flow velocity $(v_f^{ss})$, depends on the pressure distribution and its gradient along the liquid film, is given [40] as

$$v_f^{ss} \sim \left[\left(\frac{\sigma}{\mu_l}\right) R_{rod}\right]^{1/2} \qquad (D1)$$

Note that according to Eq.(D1), the steady state velocity scale depends on the rod diameter, while the rate of the liquid rise (the case of first imbibition or wetting) in the corners did not depend on the rod diameter (see Eq. 12 and Eq. 13). The corresponding viscous pressure loss ($\Delta p_{visc}$) due to the liquid flowing in the corner-induced [37] liquid films is

$$\Delta p_{visc} \sim \left(\frac{\mu_l h_Z}{a_f}\right) v_f^{ss} \qquad (D2)$$

For the case of 5 mm diameter rods and n-pentane as the evaporating liquid, we get $v_f^{ss} \sim 43$ cm/s and (maximum) $\Delta P \sim 2$ Pa when $h_Z = 55$ mm is considered. We see that for a considerable distance of ~5 cm, the viscous pressure loss is not high. For simplicity, we have assumed $a_f \sim L_X L_Y \sim 2.5$ mm$^2$. In reality, $\Delta P$ should be estimated over the entire corner-induced liquid film; the values will remain similar though. Note that the viscous pressure loss will increase if smaller rods are used; $v_f^{ss}$ will increase but the film area $(a_f)$ would decrease drastically. Thus, in these cases, $\Delta P$ must be taken into consideration during the complete design.